\documentclass[aps,pra,twocolumn,showpacs,superscriptaddress]{revtex4-2}  

\usepackage{physics}
\usepackage{graphicx}  
\usepackage{dcolumn}   
\usepackage{bm,amssymb,amsmath,dsfont}   
\usepackage[colorlinks=true,breaklinks=true,allcolors=blue]{hyperref}
\usepackage{url}
\usepackage{txfonts}
\usepackage{bbold}

\usepackage{graphics}
\usepackage{pstricks}
\usepackage{tikz}
\usepackage{color}
\usepackage{xcolor}
\graphicspath{{Figures/}}

\usepackage{tabularx}
\usepackage{multirow}
\usepackage{array}
\newcolumntype{L}[1]{>{\raggedright\let\newline\\\arraybackslash\hspace{0pt}}m{#1}}
\newcolumntype{C}[1]{>{\centering\let\newline\\\arraybackslash\hspace{0pt}}m{#1}}
\newcolumntype{R}[1]{>{\raggedleft\let\newline\\\arraybackslash\hspace{0pt}}m{#1}}

\newcommand{\eq}[1]{(\ref{eq:#1})}
\newcommand{\Eq}[1]{Eq.\,\eq{#1}}

\newcommand{\Fig}[1]{Fig.~\ref{fig:#1}}

\newcommand{\Sect}[1]{Sect.~\ref{sec:#1}}

\newcommand{\App}[1]{App.~\ref{app:#1}}

\newcommand{\Cite}[1]{Ref.~\cite{#1}}

\definecolor{applegreen}{rgb}{0.55, 0.71, 0.0}
\definecolor{byzantine}{rgb}{0.74, 0.2, 0.64}

\newcommand{\cred}[1]{{\color{red}{#1}}}
\newcommand{\cblu}[1]{{\color{blue}{#1}}}

\newcommand{\tbd}[1]{\cred{#1}}
\renewcommand{\tbd}[1]{}

\newcommand{\tbdso}[1]{\cblu{#1}}
\renewcommand{\tbdso}[1]{}

\newcommand{\IntermediateStep}[1]{&\cred{\textrm{\ (($===$ intermediate calc. steps $==>$))}}\nonumber\\ #1  
                                                           &\cred{\textrm{(($<=====================$))}}\nonumber\\}
\renewcommand{\IntermediateStep}[1]{}

\makeatletter
\let\cat@comma@active\@empty
\makeatother

\begin{document}


\title{Simulating the Berezinskii-Kosterlitz-Thouless Transition with Complex Langevin}
\author{Philipp Heinen}
\affiliation{Kirchhoff-Institut f\"ur Physik,
             Ruprecht-Karls-Universit\"at Heidelberg,
             Im~Neuenheimer~Feld~227,
             69120~Heidelberg, Germany}

\author{Thomas Gasenzer}
\affiliation{Kirchhoff-Institut f\"ur Physik,
             Ruprecht-Karls-Universit\"at Heidelberg,
             Im~Neuenheimer~Feld~227,
             69120~Heidelberg, Germany}
\affiliation{Institut f\"{u}r Theoretische Physik,
		Universit\"{a}t Heidelberg, 
		Philosophenweg 16, 
		69120 Heidelberg, Germany}


\begin{abstract}
Numerical simulations of the full quantum properties of interacting many-body systems by means of field-theoretic Monte-Carlo techniques are often limited due to a sign problem.
Here we simulate properties of a dilute two-dimensional Bose gas in the vicinity of the Berezinskii-Kosterlitz-Thouless (BKT) transition by means of the Complex Langevin (CL) algorithm, thereby extending our previous CL study of the three-dimensional Bose gas to the lower-dimensional case. 
The purpose of the paper is twofold. 
On the one hand, it adds to benchmarking of the CL method and thus contributes to further exploring the range of applicability of the method. 
With the respective results, the universality of the equation of state is recovered, as well as the long-wave-length power-law dependence of the single-particle momentum spectrum below the BKT transition. 
Analysis of the rotational part of the current density corroborates vortex unbinding in crossing the transition. 
Beyond these measures of consistency we compute quantum corrections to the critical density and chemical potential in the weakly coupled regime. 
Our results show a shift of these quantities to lower values as compared to those obtained from classical field theory. 
It points in the opposite direction as compared to the shift of the critical density found by means of the path-integral Monte-Carlo method at larger values of the coupling.
Our simulations widen the perspective for precision comparisons with experiment. 
\end{abstract}

\pacs{%
}

\maketitle

\section{Introduction}
\label{sec:Introduction}

The Berezinskii-Kosterlitz-Thouless (BKT) \cite{Kosterlitz1973,Berezinskii1971JETP...32..493B,Berezinskii1972JETP...34..610B}  transition represents a prototypical example of a topological phase transition and has been the interest of intensive experimental 
\cite{%
Bishop1980a.PRB22.5171,
Epstein1981a.PRL47.534,
Hebard1983a.PRL50.1603,
Agnolet1989a.PhysRevB.39.8934,
Trombettoni2005a.NJP7.57,
Lozovik2006a.JETPLett84.146,
Hadzibabic2006a,
Schweikhard2007a,
Clade2009a.PRL102.170401,
Sanvitto2010a.NatPhys6.527,
Yefsah2011a.PRL107.130401,
Mondal2011a.PRL107.217003,
Hung2011a.Nature470.236,
Ha2013a.PRL110.145302,
Zhao2013a.SolStatComm165.59,
Nitsche2014a.PRB90.205430,
Fletcher2015a.PRL114.255302,
Murthy2015a.PhysRevLett.115.010401,
Situ2020a.NatPhot14.517,
Hu2020a.NatComm11.1,
christodoulou2021observation,
Sunami2022a.PRL128.250402}
and theoretical studies 
\cite{
Kotsubo1984a.PRL53.691,
Minnhagen1987a.RevModPhys.59.1001,
Nomura1998a.JPhysA31.7341,
Prokofev2001a.PRL87.270402,
Prokofev2002a.PRA66.043608,
Benfatto2009a.PRB80.214506,
Holzmann2008a.PRL100.190402,
Pilati2008a.PRL100.140405,
Floerchinger2009b,
Filinov2010a.PRL105.070401,
Foster2010a,
Wang2010a.PRA81.064301,
Holzmann2010a.PRA81.043622,
Jensen2010a.PRL105.041601,
CapogrossoSansone2010a.NJP12.043010,
Plisson2011a.PRA84.061606,
Rancon2012a.PRA85.063607,
Costa2013a.PLA377.1239,
Ozawa2014a.PRL112.025302,
Kobayashi2019a.PRL123.075303,
Tononi2019a.PRL123.160403,
Kobayashi2019a.JPSJ88.094001,
Gawryluk2019a.PRA99.033615,
Julku2020a.PRB101.060505,
Zhiqiang2020a.PRB102.184504,
Miyajima2021a.PRB104.075114,
Comaron2021a.EPL133.17002c,
Tononi2022a.PRR4.013122,
Giachetti2022a.PRB106.014106},
cf., e.g., \cite{Altland2010a,Jose2013a.40yearsBKT} for theory reviews. 
Universal features of this phase transition, such as the behavior of correlation functions and the superfluid fraction at the transition, can be successfully described by the BKT renormalization  group theory \cite{Kosterlitz1973,Berezinskii1971JETP...32..493B,Berezinskii1972JETP...34..610B}. 
However, there are also non-universal characteristics that depend on the specific underlying system. 
For their quantitative determination one has to rely on Quantum Monte Carlo numerical simulations \cite{Holzmann2008a.PRL100.190402,Pilati2008a.PRL100.140405,Filinov2010a.PRL105.070401,Foster2010a} or analytical tools such as the functional renormalization group \cite{Floerchinger2009b,Rancon2012a.PRA85.063607} or the Beliaev technique \cite{CapogrossoSansone2010a.NJP12.043010}.

Here we study the example of a system undergoing a BKT transition, namely the interacting single-component Bose gas in two spatial dimensions. 
While there has been strong pro\-gress by means of the aforementioned numerical techniques, in the limit of large particle numbers, a field-theoretic approach would be desirable. 
The field-theoretic action of an equilibrium Bose gas is complex due to a Berry phase term.
When simulated by Monte Carlo techniques it is therefore subject to the so-called sign problem \cite{Henelius2000a.PRB61.1102,Lombardo2007a.MPLA22.457,Aarts2009.PhysRevLett.102.131601,Alford2010a.JPhysG37.025002,Berger2021a.PRep892.1,Troyer2005a.PRL94.170201}. 
The sign problem is absent in simulations of classical $|\psi|^4$ theory  \cite{Prokofev2001a.PRL87.270402,Prokofev2002a.PRA66.043608} and in semi-classical simulations of the Bose gas \cite{Foster2010a,Gawryluk2019a.PRA99.033615}. 
These approximative descriptions become asymptotically exact for a gas with vanishing coupling at the transition point, while for nonzero coupling and away from the transition quantum corrections  generally play a role.  
Another approach is to make use of a quantum-mechanical formulation by means of the path integral Monte Carlo (PIMC) method \cite{Holzmann2008a.PRL100.190402,Pilati2008a.PRL100.140405}, for which the sign problem is absent. 

Here we simulate the full quantum field-theoretic action from first principles by means of the complex Langevin (CL) method, which represents a generic approach for tackling the sign problem \cite{Berger2021a.PRep892.1}. 
While this method does not serve to solving the (NP-hard \cite{Troyer2005a.PRL94.170201}) sign problem for arbitrary systems and fails in certain cases, it has been shown to succeed in a broad range of problems 
\cite{%
Berges:2006xc,
Aarts2008a.JHEP2008.018,
Aarts2009.PhysRevLett.102.131601,
Aarts2010a.JHEP2010.1,
Sexty2014a.PLB729.108,
Hayata2015a.PRA92.043628,
Loheac2017a.PRD95.094502,
Rammelmuller2018a.PRL121.173001,
Nishimura2019a.JHEP2019.1,
Kogut2019a.PRD100.054512,
Ito2020a.JHEP2020.1,
Attanasio2020a.EPJA56.1,
Attanasio2020a.PRA101.033617,
Berger2021a.PRep892.1,
Attanasio:2022kkf,
boguslavski2023stabilizing} 
and can provide ab-initio simulations of the not-too-strongly coupled Bose gas \cite{Delaney2020PhysRevLett.124.070601,Heinen2022a.PhysRevA106.063308.complex}.

In \Sect{CLapproach2DBose} we briefly summarize the main aspects of the CL approach and of the near-critical characteristics of a 2D Bose gas.
Our results demonstrate that CL is a useful tool when looking for quantum corrections to non-universal features of the Bose gas that have been determined, in the regime of weak couplings considered here, by simulations of the classical model so far \cite{Prokofev2001a.PRL87.270402} (\Sect{critdens}). 
Besides that, the application of CL to the non-relativistic Bose gas is not very well established yet in general. 
We thus want to show that well-known predictions of BKT theory can be reproduced by CL simulations: the scale invariance of the equation of state near the transition (\Sect{scaleinv}), the algebraic decay of correlation functions below the phase transition (\Sect{corrfunc}) and the vortex unbinding process across the transition (\Sect{vortices}). 
These studies are to contribute in building trust in the method in itself. 
We close with a summary and outlook.

\section{Complex Langevin approach to a Bose gas}
\label{sec:CLapproach2DBose}
We begin with a brief summary of the essential aspects of the Complex Langevin (CL) method for simulating many-body quantum dynamics of a single scalar field as well as for a single-component Bose gas.
The technique and details concerning various applications has been presented in great detail in, e.g., Refs. \cite{Berger2021a.PRep892.1,Aarts2008a.JHEP2008.018,Sexty2014a.PLB729.108,Loheac2017a.PRD95.094502,Hayata2015a.PRA92.043628,Berges:2006xc}, as well as, for an interacting Bose gas in three dimensions, in Ref. \cite{Heinen2022a.PhysRevA106.063308.complex}.
In the remainder of the section we discuss essential aspects of BKT physics that we are going to probe by means of the CL approach.

\subsection{Complex Langevin equations}
\label{sec:CLmethod}
Given a quantum field theory for a single scalar field $\phi$ 
defined by an action $S[\phi]\in\mathbb{R}$, observables $\mathcal{O}(\phi)$ can be computed as 
\begin{align}
\label{eq:ExpValue}
\langle\mathcal{O}\rangle_{S}
=Z^{-1}\int \mathcal{D}\phi\,\exp(-S[\phi])\,\mathcal{O}(\phi)\,,
\end{align}
with normalization given by the partition function
\begin{align}
	\label{eq:Z}
	Z=\int \mathcal{D}\phi\,\exp(-S[\phi])
	\,.
\end{align}
Here $\int\mathcal{D}\phi$ denotes a path integral over all possible field configurations subject to the respective boundary conditions. 
In the case of thermal equilibrium and a bosonic field theory, the field $\phi(\tau,\mathbf{x})$ depends on position $\mathbf{x}$ and imaginary time $\tau\in[0,\beta]$, with inverse temperature $\beta$, and must fulfil periodic boundary conditions in imaginary time, $\phi(\tau=0,\mathbf{x})=\phi(\tau=\beta,\mathbf{x})$.

The path integral can be reformulated as a Langevin-type stochastic differential equation in a fictitious time $\vartheta$:
\begin{align}
	\label{eq:rl_eq}
	\frac{\partial\phi}{\partial\vartheta}
	=-\frac{\delta S}{\delta\phi}+\eta(\vartheta)
	\,,
\end{align}
with $\eta(\vartheta)$ being a Wiener noise subject to $\langle\eta(\vartheta)\rangle=0$ and $\langle \eta(\vartheta)\eta(\vartheta')\rangle=2\delta(\vartheta-\vartheta')$. 
Observables are evaluated as the temporal average of $\mathcal{O}(\phi(\vartheta))$ along the Langevin trajectory. 
The advantage of this formulation is that it can be easily generalized to the case of a complex action $S[\phi]\in\mathbb{C}$, while standard algorithms for evaluating \eq{ExpValue} based on importance sampling are applicable only for real actions. 
Namely, we can straightforwardly complexify $\phi\to\phi_R+\mathrm{i}\phi_I$, thereby doubling the number of (real) degrees of freedom. 
We have then to evolve twice as many Langevin equations,
\begin{align}
	\label{eq:cl_eq}
	\frac{\partial\phi_R}{\partial\vartheta}
	&=-\Re\left[\frac{\delta S}{\delta\phi}\right]+\eta_R(\vartheta)
	\,,\\
	\label{eq:cl_eq_Im}
	\frac{\partial\phi_I}{\partial\vartheta}
	&=-\Im\left[\frac{\delta S}{\delta\phi}\right]+\eta_I(\vartheta)
	\,,
\end{align}
where the real and imaginary noise, $\eta_R(\vartheta)$ and $\eta_I(\vartheta)$, must fulfill $\langle\eta_R(\vartheta)\rangle=\langle\eta_I(\vartheta)\rangle=0$ and $\langle\eta_R(\vartheta)\eta_R(\vartheta')\rangle=2N_R\delta(\vartheta-\vartheta')$, $\langle\eta_I(\vartheta)\eta_I(\vartheta')\rangle=2N_I\,\delta(\vartheta-\vartheta')$. 
$N_R$ and $N_I$ are subject to the constraint $N_R-N_I=1$ but can be chosen arbitrarily in principle. 
However, it is numerically most convenient to set $N_R=1$ and $N_I=0$ \cite{Aarts2010complex.PhysRevD.81.054508}, which we will do in the following. This  magnitude of the noise ensures, via the fluctuation-dissipation theorem, that the correct action $S$ is simulated \cite{Berger2021a.PRep892.1}.
 
Analogously to the case of a real action, observables are obtained as the temporal average of $\mathcal{O}(\phi_R+\mathrm{i}\phi_I)$ along the Langevin trajectory. 
Since the expectation value of hermitian observables must be real if the Hamiltonian is hermitian itself, the imaginary part of $\mathcal{O}(\phi_R+\mathrm{i}\phi_I)$ vanishes when averaging over the Langevin time.

\subsection{Dilute Bose gas in two spatial dimensions}
\label{sec:Model}
A one-component Bose gas in two spatial dimensions at temperature $T=1/\beta$ and chemical potential $\mu$ is described by the action
\begin{align}
	\label{eq:action}
	S[\psi,\psi^*]
	=\int \limits_0^\beta \mathrm{d}\tau\int \mathrm{d}^2x \left[\psi^*\partial_\tau\psi+\mathcal{H}(\psi,\psi^*)\right]
	\,,
\end{align}
for a complex valued field $\psi(\tau,\mathbf{x})$, with Hamiltonian density
\begin{align}
	\label{eq:Hamiltonian}
	\mathcal{H}(\psi,\psi^*)
	=\frac{1}{2m}\nabla\psi^*\cdot\nabla\psi-\mu\,|\psi|^2+\frac{g_\mathrm{B}}{2}|\psi|^4
	\,,
\end{align}
where $m$ is the mass and $g_\mathrm{B}$ the (bare) coupling constant and we have set $\hbar=k_\mathrm{B}=1$.  
As it is the case in three dimensions, the 2D bosonic quantum gas is not UV-finite, and the coupling needs to be renormalized. 

The relation between the renormalized, physical coupling $g$ and the bare coupling $g_\mathrm{B}$ can be expressed as 
\begin{align}
\label{eq:renormalization_general}
mg=\frac{4\pi\, mg_\mathrm{B}}{{4\pi}+mg_\mathrm{B}\,\log\left({\Lambda^2}/\Lambda_0^2\right)}
\,,
\end{align}
where $\Lambda$ is the momentum cutoff and $\Lambda_0$ the momentum scale at which the renormalized coupling is defined. 
Following the convention of \cite{Franca2017two} to take the square root of the particle density to be the momentum scale where the renormalized coupling is defined, $\Lambda_0=\sqrt{\rho}$, the renormalized coupling reads
\begin{align}
	\label{eq:renormalization}
	mg=\frac{4\pi\, mg_\mathrm{B}}{{4\pi}+mg_\mathrm{B}\,\log\left({\Lambda^2}/{\rho}\right)}
	\,.
\end{align}
An alternative would be to define the coupling at the scale of the healing momentum, $\Lambda_0=\sqrt{2mg\rho}$, as employed in \cite{Petrov2000bose}. The relative deviation in the renormalized coupling for the two definitions is $mg\ln(2mg)/(4\pi)$ and thus tiny in the weakly interacting regime $mg\ll 1$.
In the following, we will take the lattice cutoff to be the UV cutoff $\Lambda$ of the theory.

In experimental settings, such a system is realized by a dilute three-dimensional Bose gas highly confined in the $z$-direction, such that $mg$ is proportional to the ratio of the (three-dimensional) scattering length $a$ and the effective extension $a_z$ of the condensate in $z$-direction:
\begin{align}
	mg=\sqrt{8\pi}\,\frac{a}{a_z}
	\,.
\end{align}
In contrast to the three-dimensional case, the coupling strength is independent of the particle density and is purely determined by the dimensionless quantity $mg$.

We write the complex Bose field $\psi$ as $\psi=\varphi+\mathrm{i}\chi$, with two real fields $\varphi,\chi \in\mathbb{R}$. 
By virtue of the CL algorithm, these two real fields are again complexified, $\varphi\to\varphi_R+\mathrm{i}\varphi_I$, $\chi\to\chi_R+\mathrm{i}\chi_I$, such that it governs the dynamics of four real fields.
We quote the CL equations we solve numerically in \App{CLequations}.

\subsection{Coherence properties across the BKT transition}
\label{sec:BKT}
We briefly summarize a few of the most important characteristics of a Bose gas near the BKT transition that we will consider in the following. 
For general reviews of the theory cf., e.g., Refs.~\cite{Altland2010a,Jose2013a.40yearsBKT}. 

The phase-ordered and disordered phases, which are separated by the BKT transition, are most easily distinguished by means of the first-order spatial coherence function or equal-time two-point correlator $g(\mathbf{r})$, defined as
\begin{align}
	g(\mathbf{r})=\langle\psi^\dagger(\mathbf{r}+\mathbf{x})\psi(\mathbf{x})\rangle
	\,.
\end{align}
Considering a homogeneous situation, it approaches the total density $\rho=\langle\psi^\dagger(\mathbf{x})\psi(\mathbf{x})\rangle$ at zero distance $\mathbf{r}=0$ and, in the disordered phase above the transition temperature $T_\text{BKT}$, falls off exponentially at large distances, $g(\mathbf{r})\sim\exp\{-|\mathbf{r}|/\xi_\mathrm{c}\}$, with an in general temperature-dependent correlation length $\xi_\mathrm{c}$. 
Below the transition, $g(\mathbf{r})$ shows algebraic behavior instead,
\begin{align}
  	g(\mathbf{r})\sim \left(|\mathbf{r}|/\xi\right)^{-\alpha}
  	\,,\qquad|\mathbf{r}|\to\infty
  	\,,
  	\label{eq:grbelowBKT}
\end{align}
on length scales $|\mathbf{r}|\gg\xi$ larger than the zero-temperature healing length $\xi=(2mg\rho)^{-1/2}$.
The behavior of the momentum-space occupation number $f(\mathbf{k})$, defined as
\begin{align}
	f(\mathbf{k})=\langle\psi^\dagger(\mathbf{k})\psi(\mathbf{k})\rangle
	\,,
	\label{eq:fk}
\end{align}
correspondingly scales as
\begin{align}
	f(\mathbf{k})\sim {|\mathbf{k}|^{\alpha-2}}
	\,,\qquad |\mathbf{k}|\to 0
	\,.
	\label{eq:fkscaling}
\end{align}
The scaling exponent
\begin{align}
	\alpha=\frac{1}{\lambda_T^2\,\rho_\text{s}}
	\,
	\label{eq:scalingexponent}
\end{align}
depends on the thermal de Broglie wave length $\lambda_T=\sqrt{2\pi/mT}$ and on the superfluid density $\rho_\text{s}$, which, in $d$ spatial dimensions is defined through \cite{Pollock1987a.PRB36.8343}
\begin{align}
	\frac{\rho_\text{s}}{\rho}=1-\frac{\langle\mathbf{P}^2\rangle}{dmTN_\text{tot}}
	\label{eq:rhos}
	\,,
\end{align}
where $\mathbf{P}$ is the total momentum of the system and $N_\text{tot}$ the total particle number. 
In the thermodynamic limit, decreasing the temperature through the transition, $\rho_\text{s}$, according to Nelson and Kosterlitz \cite{Nelson1977a.PRL39.1201}, makes a universal jump from $0$ to 
\begin{align}
	\label{eq:nelson}
	\rho_{\text{s},\text{c}}
	=\frac{2mT}{\pi}
	=\frac{4}{\lambda_{T}^{2}}
	\,,
\end{align}
such that the scaling exponent approaches the critical value $\alpha=1/4$ for $T\to T_\text{BKT}^-$, while $\alpha$ goes to $0$ for $T\to 0$.
Hence, at non-zero temperatures, near the BKT transition, the superfluid density depends on temperature and thus, non-linear couplings between the modes become important below the then $T$-dependent healing-length scale $k_\text{c}\sim\sqrt{mg\rho_\mathrm{s}}\sim m\sqrt{gT}\sim \sqrt{mg}\, k_{T}$, with thermal momentum $k_{T}\sim\sqrt{mT}$.
For momenta $k_\mathrm{c}\ll k\ll k_{T}$, the occupation number takes the Rayleigh-Jeans form $f(\mathbf{k})=2mT/|\mathbf{k}|^2$. 

Whereas the \textit{superfluid} density \eq{nelson} takes a universal value at the transition, the \textit{total} density $\rho$ does not and is dependent on the microscopic details of the system instead.
It has been predicted in \cite{Prokofev2001a.PRL87.270402}, on the basis of the observation that the major contribution to the total density comes from the range of momenta $k_\text{c}\lesssim k\lesssim k_\text{UV}\sim k_{T}$, that, for $mg\ll1$, the critical density and chemical potential behave as
\begin{align}
	\label{eq:critical_dense}
	\rho_\text{c}
	&=\frac{1}{\lambda_T^2}\ln\left(\frac{\zeta_{\rho}}{mg}\right)
	\,,
	\\
	\label{eq:critical_mu}
	\mu_\text{c}
	&=\frac{2g}{\lambda_T^2}\ln\left(\frac{\zeta_{\mu}}{mg}\right)
	\,.
\end{align}
The numerical constants were given as $\zeta_{\rho}=380\pm3$ and $\zeta_{\mu}=13.2\pm0.4$ \footnote{%
In \cite {Prokofev2001a.PRL87.270402}, $\protect \zeta_{\rho}$ and $\protect \zeta_{\mu }$ were denoted as $\xi $ and $\xi _{\mu }$, respectively.}. 
As these were obtained from simulations of the classical field theory, which become exact in the limit $mg\to 0$, it will be interesting to check this prediction against the simulations of the full quantum model at nonzero $mg$. 

From the above results, one finds that the weakly coupled 2D Bose gas exhibits universal behavior across the BKT transition:
As was argued in \Cite{Prokofev2001a.PRL87.270402}, both, $\rho_\text{c}$ and $\mu_\text{c}$ are non-universal quantities, that depend on the ultraviolet (UV) cutoff scale $k_\text{UV}$, through the relation $\zeta_{\mu}/mg\equiv Ck_\text{UV}^{2}/k_\text{c}^{2}$, etc., while the cutoff dependence cancels out in their quotient $\zeta_{\rho}/\zeta_{\mu}$. 
As a result, also $\mu_\text{c}-2g\rho_\text{c}$ is cutoff-independent and thus universal.
Hence, it was proposed that this remains so in the vicinity of the critical point.
Specifically, this means that, subtracting the chemical potential from the mean-field-type expression $2g\rho$, that one has $(2g\rho-\mu)/(mgT)=\theta(X)$, and thus that the equation of state can be written in terms of a universal function $\theta(X)$ that depends on the dimensionless tuning parameter $X=(\mu-\mu_\text{c})/mgT$ only, valid within the so-called fluctuation region, where $X$ varies on the order of unity.
Evaluating the difference of this relation to the critical one leads to the equation of state in the form 
\begin{align}
   \rho(\mu)-\rho_\text{c}
   =\frac{1}{\lambda_T^2}F\left(\frac{\mu-\mu_\text{c}}{mgT}\right)
\label{eq:eosuniversal}
\,,
\end{align}
where the universal function $F(X)$ is related to $\theta(X)$ by
\begin{align}
\label{eq:EOS_mean_field}
   F(X)=\pi\left[\theta(X)+X-\frac{1}{\pi}\ln\left(\frac{\zeta_{\rho}}{\zeta_{\mu}}\right)\right]\,,
\end{align}
In the Bogoliubov mean-field limit, $X\to\infty$, the function $\theta$ asymptotically obeys 
\begin{align}
   \theta(X)-\frac{1}{\pi}\ln\left[\theta(X)\right]
   \to X+\frac{1}{\pi}\ln(2\zeta_{\mu})
   \label{eq:thetaMF}
   \,,
\end{align}
whereas for $X\to-\infty$, a Hartree-Fock approximation yields
\begin{align}
\theta(X)+\frac{1}{\pi}\ln\left[\theta(X)\right]
\to |X|-\frac{1}{\pi}\ln(\zeta_{\mu})
\label{eq:thetaMF2}
\,,
\end{align}
and the resulting universal scaling form \eq{eosuniversal} of the equation of state was confirmed in this limit by means of classical Monte Carlo simulations \cite{Prokofev2002a.PRA66.043608} for different values of the coupling $mg$.

Eventually, also the superfluid density, for chemical potentials below the BKT transition and in its vicinity of the transition, obeys a functional form generalising the Nelson-Kosterlitz result \eq{nelson} \cite{Prokofev2002a.PRA66.043608},
\begin{align}
	\label{eq:nelsoninfluctreg}
	\rho_{\text{s}}
	=\frac{2mT}{\pi}\,f(X)
	\,,
\end{align}
with a universal function $f(X)$ obeying $f(X\to0^{+})=1$ and $f(X<0)\equiv0$.

\subsection{Topological characteristics across the BKT transition}
\label{sec:TopoBKT}
A further important characteristic of the BKT transition is topological in nature: 
Whereas above $T_\text{BKT}$ one considers the gas to be dominated by free quantum vortices, which deteriorate the phase coherence over short distances, these mutually annihilate as pairs of vortices and anti-vortices and thus disappear below the transition, giving rise to an algebraically slow decoherence \eq{grbelowBKT}.

To study topological properties, we evaluate the current density, defined by
\begin{align}
	\mathbf{j}
	=\frac{1}{2m\mathrm{i}}\left[\psi^*\nabla\psi-\psi\nabla\psi^*\right]
	\label{eq:current}
	\,.
\end{align}
According to the Helmholtz theorem, $\mathbf{j}$ can be decomposed into an irrotational and rotational part 
\begin{align}
	\mathbf{j}
	=\mathbf{j}_\text{irr}+\mathbf{j}_\text{rot}
	\,.
\end{align}

Under the simplifying assumptions that density fluctuations are negligible and free vortices are uncorrelated, the local density of free (unbound vortices) can be estimated in terms of the mean of the rotational part of the current squared as
\begin{align}
	\label{eq:defrhovfree}
	\rho_{\mathrm{v,free}}(\mathbf{x}) 
	&\simeq \frac{m^2}{2\pi \ln(L/\xi_\text{h})\langle\rho\rangle^2}\,|\mathbf{j}_\text{rot}(\mathbf{x})|^2
	\,,
\end{align}
where $L$ is the system size and $\xi_\mathrm{h}=(2m\mu)^{-1/2}$ the healing length. 
A decrease of $\rho_{\mathrm{v,free}}$ across the transition reflects the vortex-anti-vortex recombination process, when going over from the disordered above to the ordered phase below the transition.

\subsection{\label{sec:finitesize}Finite-size scaling}
Numerical simulations must necessarily be performed in a finite volume. For obtaining properties in the thermodynamic limit, some extrapolation is necessary since finite-size corrections vanish only logarithmically in a two-dimensional Bose gas. 
BKT renormalization group (RG) theory allows determining the finite-size scaling of the superfluid density. 
On the basis of general scaling arguments, $\rho_\mathrm{s}(L)$, close to criticality, can be written as a singular function of $L$ times a non-singular function of $L/\xi_T$, where $\xi_{T}\sim\xi_\mathrm{c}{\exp}[a(T-T_\mathrm{BKT})^{-1/2}]$ is the (at criticality diverging) correlation length \cite{Sandvik2010a.AIPCP1297.135} and $\xi_\mathrm{c}\sim 1/k_\mathrm{c}(T_\mathrm{BKT})\sim(m\sqrt{gT_\mathrm{BKT}})^{-1}$ is a microscopic scale on the order of the healing length near the critical point. 
For the exact functional form, one needs to solve the BKT RG equations.
Keeping the temperature $T$ fixed, one obtains, in leading order, the finite-size correction to \eq{nelson} at the infinite-size critical chemical potential $\mu_\text{c}(L\to\infty)$ \cite{Weber1988a.PRB37.5986},
\begin{align}
\label{eq:rhos_LO}
   \rho_\mathrm{s}^\text{LO}(L)
   =\frac{2mT}{\pi}\Bigg(1&+\frac{1}{2\ln (L/\xi_\mathrm{c})+C}\Bigg)
   \,,
\end{align}
with some constant $C$, which we have explicitly written, other than usually found in the literature, in terms of the near-critical healing-length scale $\xi_\text{c}$ \footnote{%
When integrating the BKT RG equations from the microscopic scale $\xi_\mathrm{c}$ to the volume scale $L$, the constant $C$ defines the initial value of the $L$-dependent function $f_{L}$, \Eq{nelsoninfluctreg}, at the microscopic scale $\xi_{c}$.
For example, in the approximation chosen in \eq{rhos_LO}, it is the inverse of the deviation of $f_{\xi_\mathrm{c}}$ from unity, $C=(f_{\xi_\mathrm{c}}-1)^{-1}$.
},
which later-on, for a fixed temperature, determines the scaling in terms of $mg$.
Higher-order expressions are also available \cite{Hsieh2013a.JStatMechTE2013.P09001}. 
For example, to next-to-leading order one has
\begin{align}
\label{eq:rhos_NLO}
   \rho_\mathrm{s}^\text{NLO}(L)
   =\frac{2mT}{\pi}\Bigg(1&+\frac{1}{2\ln (L/\xi_\mathrm{c})+C+\ln\left[C/2+\ln(L/\xi_\mathrm{c})\right]}\Bigg)
   \,.
\end{align}
Instead of extrapolating the superfluid density to the infinite-size limit for each single parameter set, we will follow the somewhat different procedure described in \cite{Prokofev2001a.PRL87.270402}. 
Namely, one defines a finite-size critical potential $\mu_\mathrm{c}(L)$ by demanding that the Nelson-Kosterlitz criterion \cite{Nelson1977a.PRL39.1201}, \Eq{nelson}, be fulfilled at $\mu_\mathrm{c}(L)$ in the finite-size system \footnote{Note that different criteria could be used to define the critical point such as the power-law exponent of the correlation function to take the value $\alpha=1/4$.}.
Using then the universal form \eq{nelsoninfluctreg} away from criticality, with an $L$-dependent function $f_{L}(X)$, one finds  that the Kosterlitz-Thouless RG equations, in the $L\to\infty$ limit, cause $f=f_{L\to\infty}$ to obey the relation $f^{-1}+\ln f=1+\kappa'X$ near $X=0$, with a numerical constant $\kappa'=0.61(1)$  \cite{Prokofev2002a.PRA66.043608}.
Expanding $f$ about unity this yields that, at $\mu$ closely above $\mu_\mathrm{c}(L)$, the superfluid density obeys
\begin{align}
  \frac{\rho_\mathrm{s}\pi}{2mT}-1
  =f\left(\frac{\mu-\mu_\mathrm{c}(L)}{mgT}\right)-1
  \simeq \sqrt{\frac{2\kappa'[\mu-\mu_\mathrm{c}(L)]}{mgT}}
  \,.
\end{align}
Combining this with the leading-order approximation \eq{rhos_LO}, the finite-size scaling form of the difference between $\mu_\mathrm{c}(L\to\infty)$ and $\mu_\mathrm{c}(L)$ results as 
\begin{align}
  \mu_\mathrm{c}(L\to\infty)-\mu_\mathrm{c}(L)
  \overset{\text{LO}}{\simeq} \frac{mgT}{2\kappa'\left[2\ln (L/\xi_\mathrm{c})+C\right]^2}
  \,.
\end{align}
Finally, since the function $F(X)$ defined in \Eq{eosuniversal} is, in leading order, linear in $X$ close to criticality ($X=0$), one finds that $\rho-\rho_\mathrm{c}(L)\sim [\mu-\mu_\mathrm{c}(L)]/g$.
Inserting $1/\xi_\mathrm{c}= m\sqrt{gT}$ at the fixed near-critical temperature $T$, one obtains the finite-size scaling of the critical density $\rho_\text{c}(L)$,
\begin{align}
\label{eq:finitesize}
\rho_\text{c}^\text{LO}(L)=\rho_\text{c}(L\to\infty)-\frac{AmT}{\ln^2\left(BLm\sqrt{gT}\right)}
\,,
\end{align}
with dimensionless constants $A$ and $B$, as introduced by \cite{Prokofev2001a.PRL87.270402}.
Analogously, the next-to-leading-order expression \eq{rhos_NLO} yields
\begin{align}
\label{eq:finitesizeNLO}
  \rho_\text{c}^\text{NLO}(L)
  &=\rho_\text{c}(L\to\infty)
    \nonumber\\
  &\ \ \ -\frac{AmT}{\left[\ln\left(BLm\sqrt{gT}\right)+\frac12\ln\ln\left(BLm\sqrt{gT}\right)\right]^2}
\,.
\end{align}
For the extraction of the critical chemical potential from finite-size data, similar relations apply. The leading order expression reads
\begin{align}
\label{eq:finitesizemu}
\mu_\text{c}^\text{LO}(L)=\mu_\text{c}(L\to\infty)-\frac{A_{\mu}mgT}{\ln^2\left(B_{\mu}Lm\sqrt{gT}\right)}
\,,
\end{align}
and the next-to-leading-order expression is given by
\begin{align}
\label{eq:finitesizemuNLO}
&\mu_\text{c}^\text{NLO}(L)
=\mu_\text{c}(L\to\infty)
\nonumber\\
&\qquad\qquad -\frac{A_\mu mgT}{\left[\ln\left(B_\mu Lm\sqrt{gT}\right)+\frac12\ln\ln\left(B_\mu Lm\sqrt{gT}\right)\right]^2}
\,.
\end{align}

\section{Results of the CL simulations}
\label{sec:results}
In the following we present the results of our CL simulations of properties of a dilute two-dimensional Bose gas close to the BKT transition. 
We begin with a determination of the critical density $\rho_\text{c}$ as a function of $mg$ in order to check the validity of the prediction \eq{critical_dense} in the quantum regime. 
Subsequently, we evaluate the equation of state and the single-particle momentum spectrum and their scaling properties near criticality, and our study closes with an evaluation of the vorticity across the BKT transition. 

We express all quantities in units of the spatial lattice constant $a_\text{s}$ (not to be confused with the $s$-wave scattering length $a$) and choose $2ma_\mathrm{s}=1$. 
Unless specified differently, our simulations were done on a grid with $256^2$ spatial $\times\, 32$ (simulations in \Sect{critdens}) or $\times\, 16$ (elsewhere) lattice points in the imaginary-time direction, with periodic boundary conditions imposed in both the temporal and spatial directions. 
We set the temporal lattice spacing to $a_\text{t}=0.05\,a_\text{s}$, resulting in a temperature $T=1.25\,a_\text{s}^{-1}$ and a thermal wave length $\lambda_T=\sqrt{2\pi/mT}=3.17\,a_\text{s}$. 
For the simulation of the Langevin equation we use a simple Euler-Maruyama scheme with time step chosen to be $\Delta\vartheta=5\cdot 10^{-3}\,a_\text{s}^{-3}$. 
While this first-order scheme is employed in the majority of CL works, also uses of higher-order \cite{Aarts2012complex} and implicit schemes \cite{Alvestad2021a.JHEP2021.8.1} have been reported.

Parameters are chosen such that the effects of spatial and imaginary-time discretization are typically smaller than the statistical errors of the CL simulations. 
The former can be estimated by comparing the discretized and continuous versions of the non-interacting gas. 
Discretization errors only affect the high-momentum modes, where the system is effectively non-interacting. 
For $N_\tau=32$ and $\lambda_T=\sqrt{2\pi/mT}=3.17\,a_\text{s}$, this yields, as estimates for the errors on the density and the superfluid density, $\lambda_T^2\delta \rho\equiv\lambda_T^2(\rho^\text{cont}-\rho^\text{latt})=4\cdot 10^{-5}$ and $\lambda_T^2\delta \rho_\text{s}/4\equiv\lambda_T^2(\rho^\text{cont}_\text{s}-\rho^\text{latt}_{s})/4=-6\cdot 10^{-3}$, which are negligible in comparison with typical statistical errors.

Moreover, typical densities result on the order of $\rho\sim 1\,a_\text{s}^{-2}$. 
Inserting the lattice cutoff for the UV cutoff, $\Lambda=\pi/a_\text{s}$, \Eq{renormalization} yields a relative deviation between the bare coupling $g_\mathrm{B}$ and the renormalized coupling $g$ in the range of $\sim2\cdot 10^{-3}$ to $\sim4\cdot 10^{-2}$ for the coupling strengths considered in the following ($mg_\mathrm{B}=0.0125\dots 0.2$). 
Since these shifts are rather tiny, we can, in the following, avoid an explicit distinction between $g$ and $g_\mathrm{B}$ and choose the bare coupling that enters the simulations as equal to the renormalized coupling $g$.
\begin{figure}
	\includegraphics[width=0.9\columnwidth]{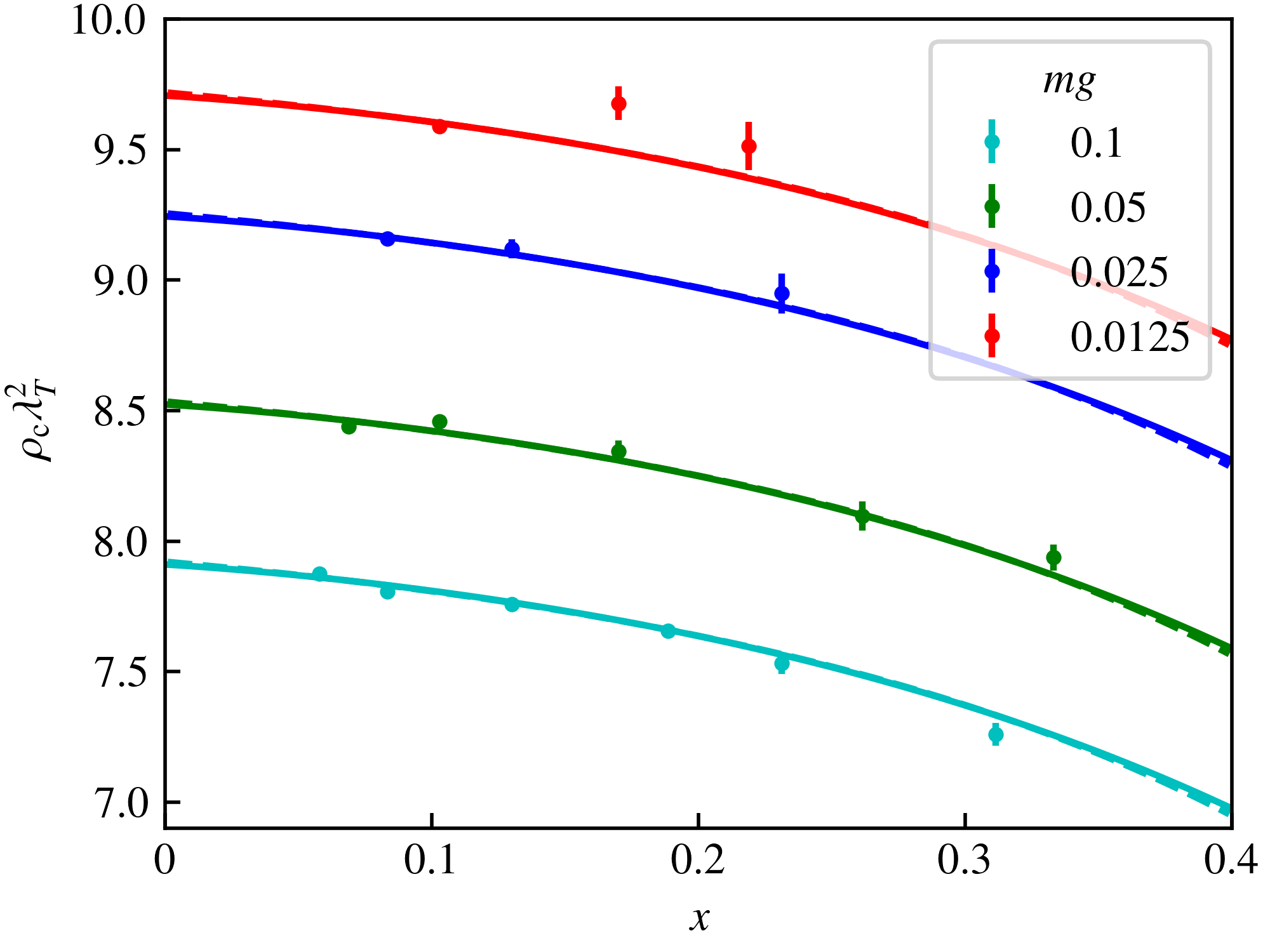}
	\includegraphics[width=0.9\columnwidth]{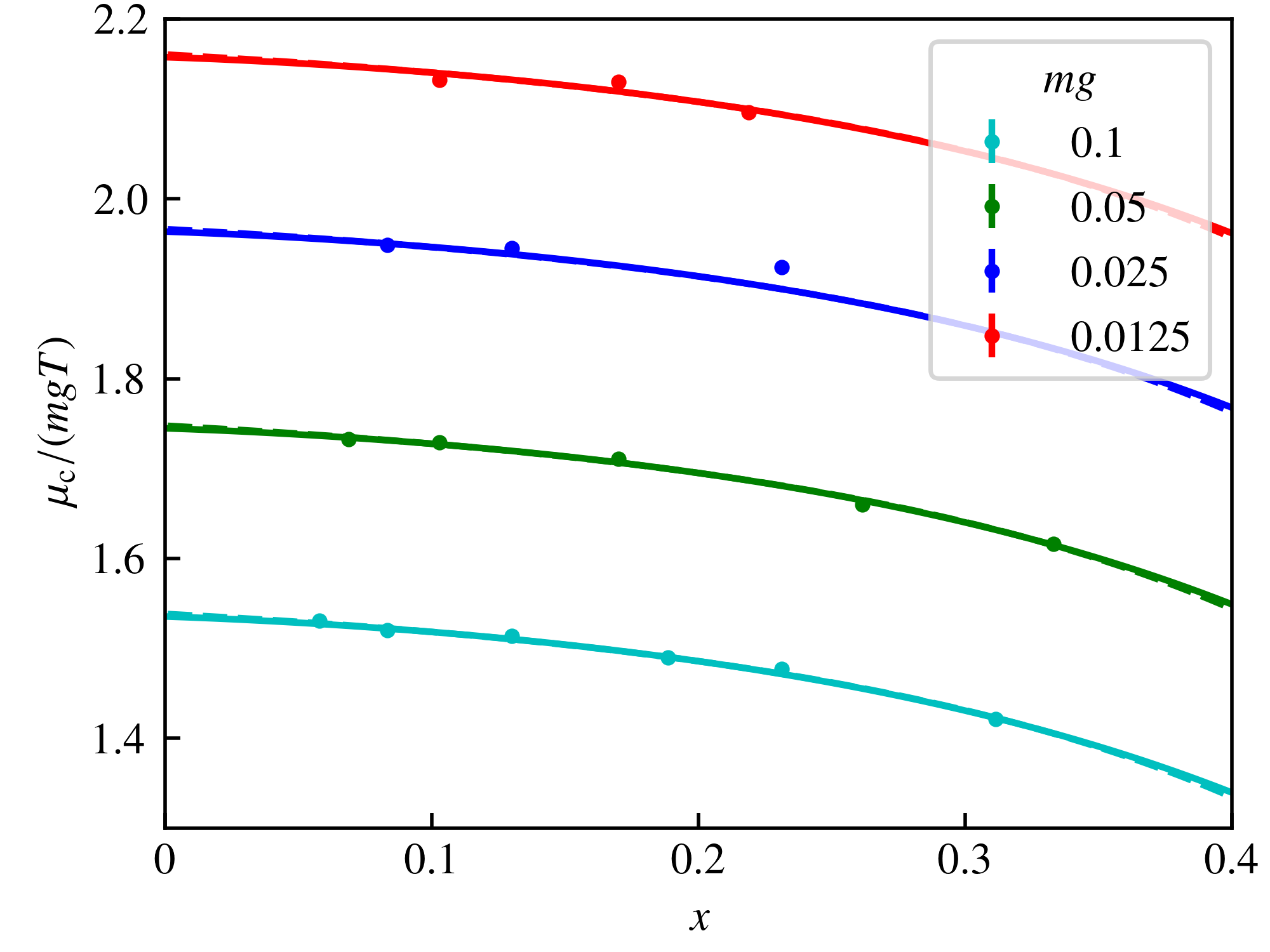}
	\caption{Critical density $\rho_\text{c}$ (upper panel) and chemical potential $\mu_\text{c}$ (lower) for various coupling strengths $mg$ as a function of the dimensionless parameter $x=\ln^{-2}\left(Lm\sqrt{gT}\right)$. Solid lines represent a fit of \eq{finitesize} and \eq{finitesizemu} to the data points. Extrapolating them for $x \to 0$ we can recover the critical density and chemical potential in the infinite-volume limit. Fits of the higher-order formula \eq{finitesizeNLO} is shown as dashed lines, which hardly deviates from the leading-order expression. 
	Where no error bars are seen, they are shorter than the width of the data points.
	}
	\label{fig:finitesize}
\end{figure}

Due to a fundamental sign problem, CL simulations are limited to a region below some maximum coupling strengths $mg$ beyond which the non-linearity in the evolution equations causes runaway trajectories which are impossible to average over in a reliable manner. 
The value of the coupling beyond which such effects dominate also depends on the proximity of the system to the BKT transition. 
Whereas far away from the transition we could simulate up to $mg\lesssim 0.4$, closer to criticality runaway trajectories frequently occur already for $mg\sim 0.2$.
These are no sharp thresholds and it is possible that larger coupling strengths can be reached by employing regulator techniques or adaptive step sizes \cite{Aarts:2010vk,Seiler2013a.PLB723.213,Attanasio2019a.EPJC79.1,Alvestad2021a.JHEP2021.8.1}. 
In probing the BKT transition we restrict ourselves to the regime $mg\leq0.1$, where runaway trajectories have been found to be absent and (polynomial) observables appear to converge well.
When studying the equation of state away from the transition, we also include a coupling $mg=0.2$. 
We have checked from our simulations that the distribution function of the magnitude of the drift, $\delta S/\delta\psi$, has compact support in this regime, which has been identified as a necessary and sufficient criterion for the correctness of the complex Langevin method \cite{Nagata2016a.PRD94.114515}.

\subsection{\label{sec:critdens}Critical density and chemical potential}
%
\begin{figure}
	\includegraphics[width=0.9\columnwidth]{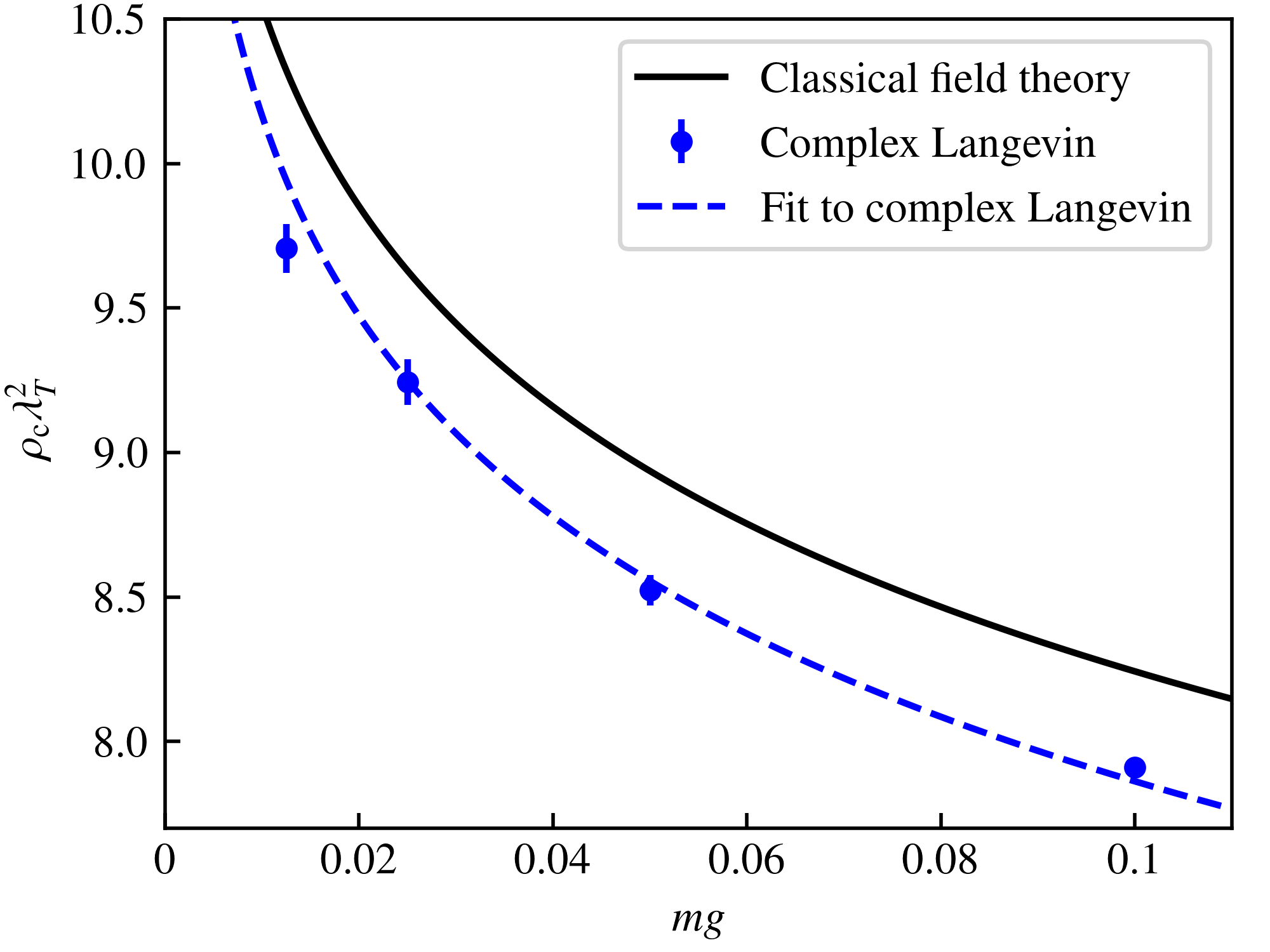}
	\includegraphics[width=0.9\columnwidth]{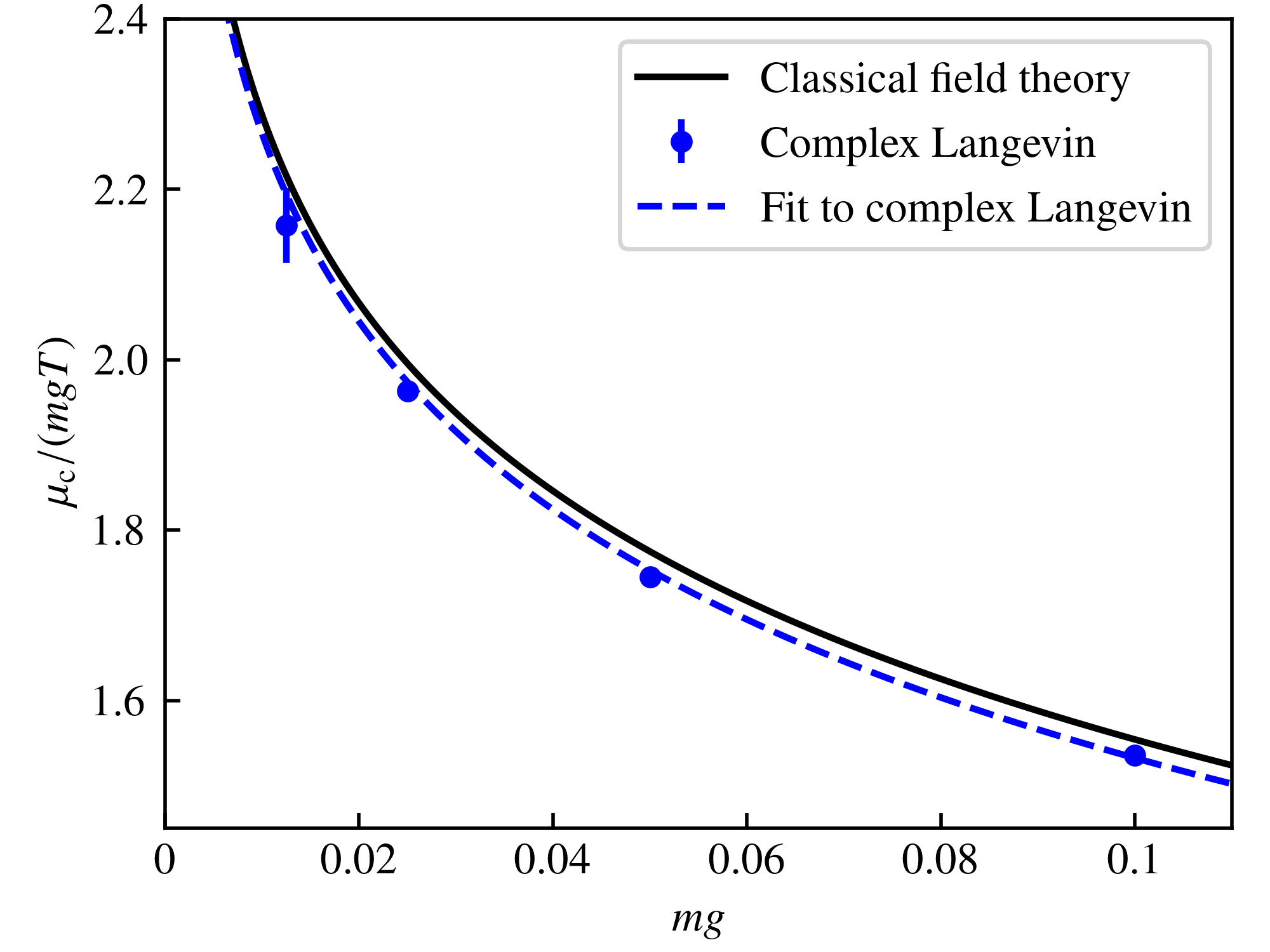}
	\caption{Critical density $\rho_\text{c}$ (upper panel) and chemical potential $\mu_\text{c}$ (lower) in the infinite-volume limit as functions of the coupling $mg$ computed with the complex Langevin simulation of the full quantum model (blue points), within a regime of couplings for which runaway processes are absent. The results are compared to the predictions \eq{critical_dense} with constant $\zeta_{\rho}=380$ and \eq{critical_mu} with $\zeta_{\mu}=13.2$ as obtained from the classical simulation of \cite{Prokofev2001a.PRL87.270402} (black lines). The critical quantities appear to be shifted downwards in the full quantum simulation by about $4\%$ ($2\%$). The blue dashed lines represent \eq{critical_dense} and \eq{critical_mu} with constant $\zeta_{\rho}=260\pm12$ and \eq{critical_mu} with $\zeta_{\mu}=12.3\pm0.1$, as obtained from fits to our data.}
	\label{fig:rhoc}
\end{figure}

We start by extracting, from our complex-Langevin data, the critical density and chemical potential in the thermodynamic limit as functions of the coupling strength, in the weak-coupling regime.
To this end, we need to perform a finite-size scaling analysis as summarized in \Sect{finitesize}.
Our results can be compared with those from classical simulations, and, within limits, with  quantum, path-integral Monte Carlo as well as experimental results.

Making use of the definition \eq{rhos} we extract the superfluid density $\rho_\text{s}$, see \App{supdens} for details of the numerical evaluation of $\langle\mathbf{P}^2\rangle$.
We then determine the critical chemical potential at the transition point by matching the result of \eq{rhos} to the Nelson-Kosterlitz criterion \eq{nelson}.
While our temporal lattice spacing fixes the temperature $T=1.25\,a_\text{s}^{-1}$, we vary the chemical potential $\mu$, whereby both the total and the superfluid density change. 
As described in more detail in \App{regfals}, we apply the secant algorithm to tune $\mu$ to the point, where the superfluid density equals the value given by \eq{nelson}.

As outlined in \Sect{finitesize}, one must expect the corresponding critical (total) density $\rho_\mathrm{c}$ to be subject to finite-size effects.
We make use of the leading-order and next-to-leading order finite-size scaling forms \eq{finitesize} and \eq{finitesizeNLO}, respectively, which we found to be indistinguishable within our statistical errors.  

We computed the finite-size critical densities for four coupling strengths, $mg=0.1$, $0.05$, $0.025$, and $0.0125$, for several lattice sizes $L/a_\text{s}\in[24,\dots,256]$. 
These are shown in \Fig{finitesize} vs.~the dimensionless parameter $x\equiv\ln^{-2}\left(Lm\sqrt{gT}\right)$. Error bars are obtained from the statistical variance of $12$ statistically independent runs.
Fitting \Eq{finitesize} to our data points for all couplings and lattice sizes simultaneously we obtain $A=0.089\pm  0.015$ and $B=0.446\pm 0.048$. The parameters $A$ and $B$ yield the $x$-dependencies of the critical density shown as solid lines in the upper panel of \Fig{finitesize}.
By extrapolating these curves to $x=0$ we obtain our prediction for the infinite-size critical densities $\rho_\text{c}(mg)$ for the four different couplings $mg$. 
Our results of this are depicted in the upper panel of \Fig{rhoc}, in comparison with the functional dependence \eq{critical_dense} obtained in \cite{Prokofev2001a.PRL87.270402} from classical field theory simulations. 
Errors are the standard fit errors on the four fitting parameters $\rho_\text{c}(mg)$, which are defined as the width of the likelihood function in the direction of the respective parameter. 
This means, using a likelihood function $L(\mathbf{p})$ with parameter vector $\mathbf{p}$ and optimal parameters $\bar{\mathbf{p}}$, we define the error in $p_i$ as $\Delta p_i=(-L/\partial_i^2L)^{1/2}\,\big|_{\mathbf{p}=\bar{\mathbf{p}}}$.

One observes that the critical densities are shifted by around $4\%$ downwards as compared to the result from the classical simulations. 
If one assumes that the universal functional form of the critical density \eq{critical_dense} still holds in the full quantum model at (small) finite $mg$, albeit with a different non-universal constant $\zeta_{\rho}$, one may fit \eq{critical_dense} to our data, yielding
\begin{align}
	\zeta_{\rho}
	&=260\pm 12
	\label{eq:zetarho}
	\,,
\end{align}
as compared with $\zeta_{\rho}=380\pm3$ obtained in \cite{Prokofev2001a.PRL87.270402}. 
Note that the relative statistical error on $\zeta_{\rho}$ is larger than the relative shift between $g_\mathrm{B}$ and $g$ even for the largest considered coupling, $(g_\mathrm{B}-g)/g\sim 1.6\%$ for $mg_\mathrm{B}=0.1$, which justifies our approximation $g_\mathrm{B}\approx g$.

The above result for the critical density may be compared with the experimental findings of \cite{Hung2011a.Nature470.236}, which gave a downward shift of $\sim 10\%$ in comparison to \eq{critical_dense} with $\zeta_{\rho}=380\pm3$.  
The experiment was conducted using an inhomogenous trapping potential, at couplings $mg=0.05,0.13,0.19,0.26$. Employing a local density approximation, the equation of state was determined from the density profile, from which in turn the authors extracted the critical densities. 
The effect of the local density approximation may need to be taken into account when comparing the by about a factor of two larger deviation with the present results. 

A different experimental approach was pursued in Ref. \cite{christodoulou2021observation}, where the authors employed the vanishing of the second-sound resonance as a criterion for the transition point in a uniform Bose gas. 
Similarly to \cite{Hung2011a.Nature470.236}, the critical temperature was found to be shifted by $\sim10\%$ upwards, i.e.~the critical density by $\sim10\%$ downwards in comparison to the classical field theory result, though the statistical error is on the same order of magnitude as the shift itself. 
Additionally, the authors point to a possible bias due to finite-size effects. 
The results of further experiments \cite{Ha2013a.PRL110.145302,Fletcher2015a.PRL114.255302} are consistent, within their error bars, with either of the present results and the predictions of \cite{Prokofev2001a.PRL87.270402}.

In contrast to our findings, the PIMC simulations of \cite{Pilati2008a.PRL100.140405} gave a shift of the critical temperature to lower values, i.e., of the critical density $\rho_\mathrm{c}$ to values higher than the ones predicted in  \cite{Prokofev2001a.PRL87.270402}. 
These simulations were performed, however, in the strong-coupling regime $mg> 1$ ($mg=1.36,2.73,5.46$) that we did not find being accessible to CL. 
As described in the next section, we can infer a value for the critical density at $mg=0.2$ from a scaling analysis of the equation of state.

For comparison, we also performed simulations of the classical field theory ourselves, as we can straightforwardly turn off quantum effects in our simulation by setting $N_\tau=1$. The results are presented in \App{classsim}. 

We performed an analogous finite-size analysis for the critical chemical potential, in which we fitted our data by means of the leading-order expression \eq{finitesizemu}.
This gave $A_{\mu}=0.087 \pm 0.003$ and $B_{\mu}=0.4 \pm 0.007$, and the corresponding fits are shown in the lower panel of \Fig{finitesize}.
(Note that one should have $B=B_{\mu}$, which is confirmed within errors.
The same does not apply to $A$ and $A_{\mu}$.)
The corresponding infinite-size $\mu_\text{c}$ is shown, again vs.~the coupling $mg$ and in comparison with the results obtained from the classical simulations of \cite{Prokofev2001a.PRL87.270402}, in the lower panel of \Fig{rhoc}.
Assuming again the functional form of the critical chemical potential \eq{critical_mu} to hold in the full quantum model at small  $mg$, a fit of \eq{critical_mu} to our data yields
\begin{align}
	\zeta_{\mu}
	&=12.3\pm 0.1
	\label{eq:zetamu}
	\,,
\end{align}
as compared with $\zeta_{\mu}=13.2\pm0.4$ obtained in \cite{Prokofev2001a.PRL87.270402}.

\subsection{\label{sec:scaleinv}Equation of state and scale invariance}
\begin{figure}
	\includegraphics[width=0.9\columnwidth]{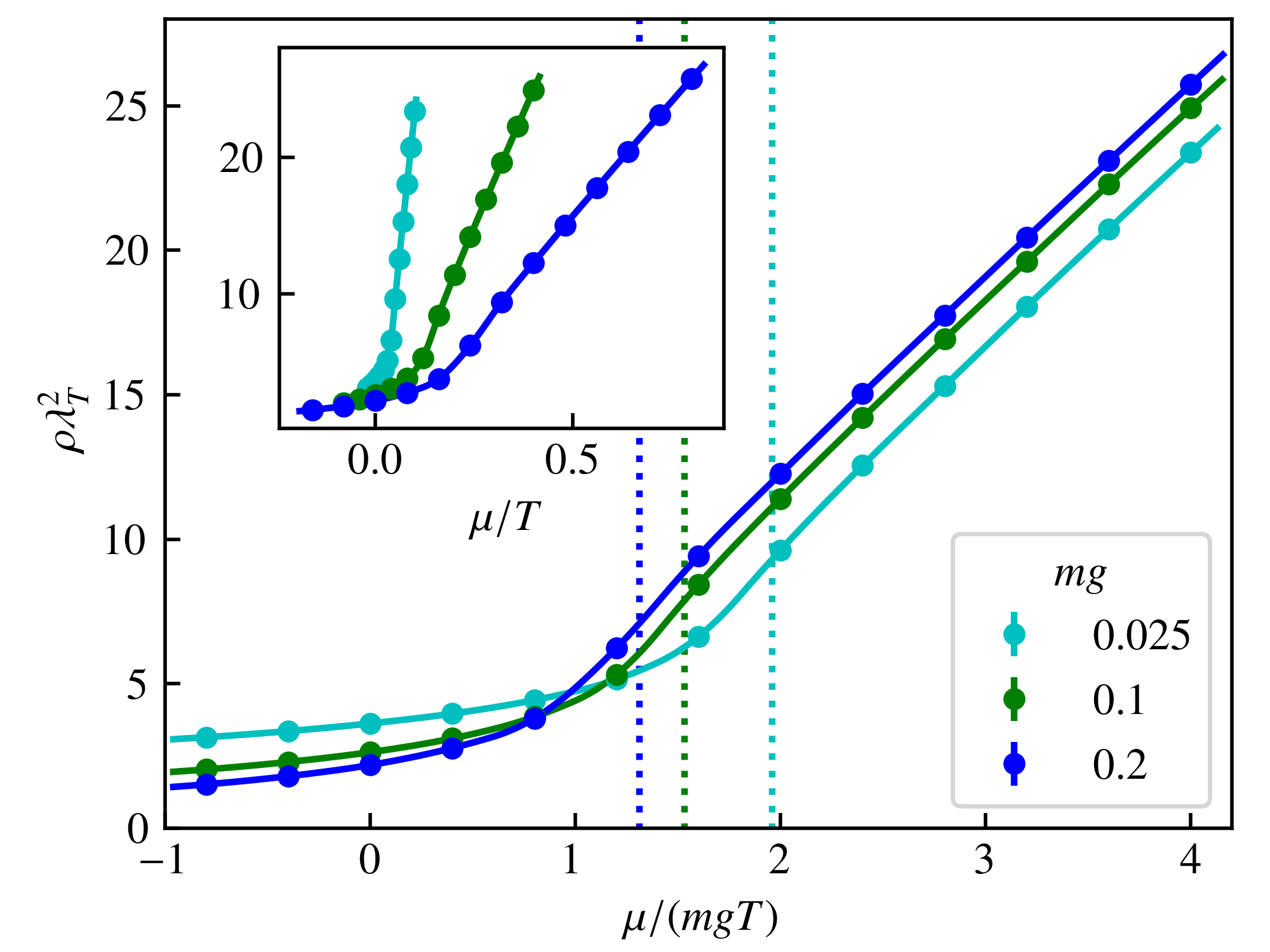}
	\includegraphics[width=0.9\columnwidth]{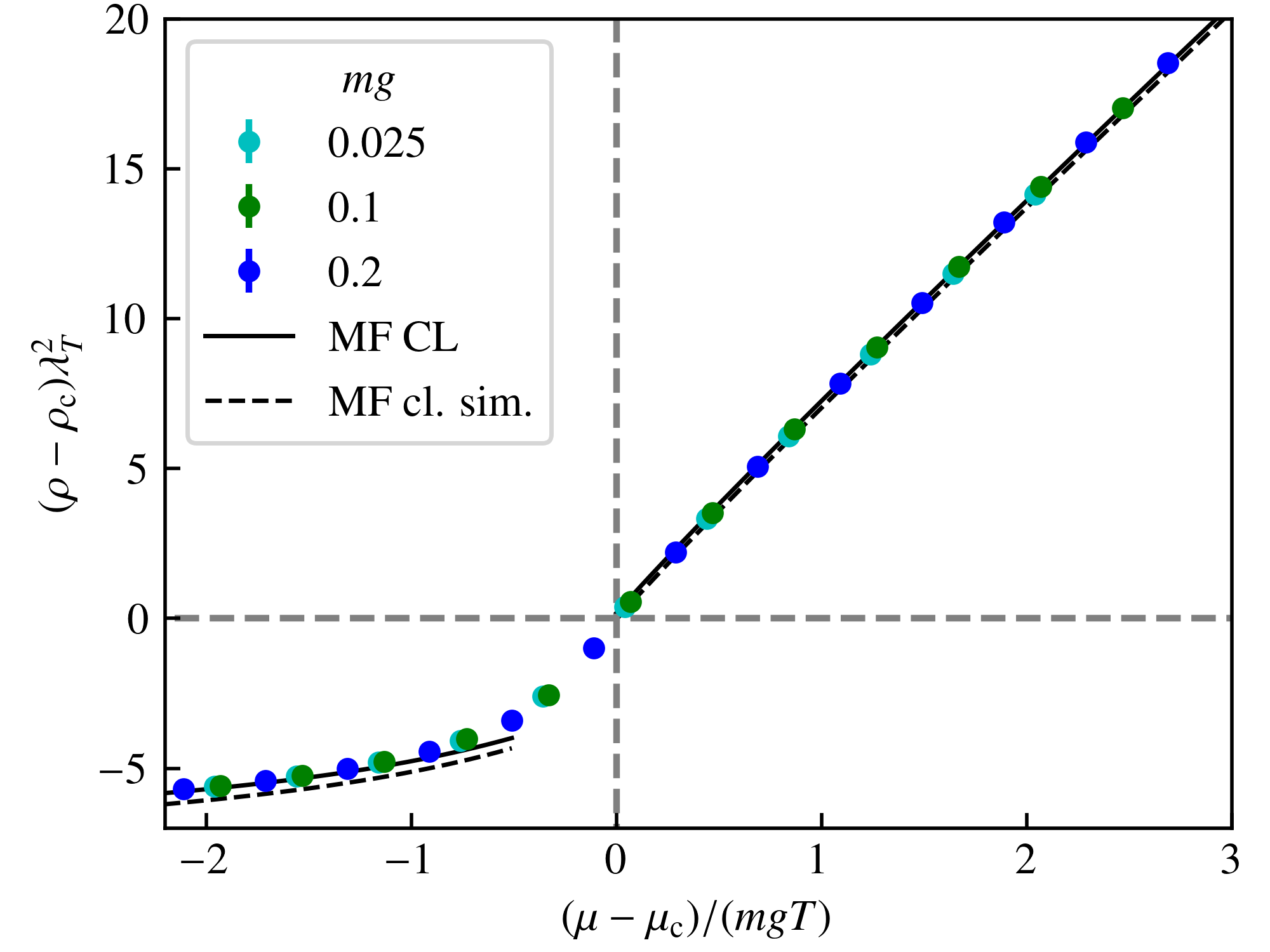}
	\caption{Equation of state, relating density $\rho$ and chemical potential $\mu$, for three values of the coupling $mg$. 
	The curves connecting the data points are obtained by spline interpolation. 
	Error bars are too small to be visible. 
	The inset of the upper panel shows the densities as functions of $\mu/T$. 
	In the main part of the upper panel, their arguments are rescaled with the coupling $mg$. The critical chemical potentials are shown as dotted vertical lines in the respective color.
	In the lower panel, the resulting curves are shifted by the critical chemical potentials and densities, cf.~\Eq{eosuniversal}. 
	The curves collapse for all three couplings considered, giving evidence of the universality of the 2D Bose gas near the BKT transition. 
	For the critical chemical potentials and densities $\rho_\text{c}$ and $\mu_\text{c}$ we took the result from  \Sect{critdens} for $mg=0.1$, whereas $\rho_\text{c}$ and $\mu_\text{c}$ for $mg=0.025$ and $mg=0.2$ are obtained from the differences $\mu_\text{c}(mg=0.025)-\mu_\text{c}(mg=0.1)$ and $\mu_\text{c}(mg=0.2)-\mu_\text{c}(mg=0.1)$ as well as $\rho_\text{c}(mg=0.025)-\rho_\text{c}(mg=0.1)$ and $\rho_\text{c}(mg=0.2)-\rho_\text{c}(mg=0.1)$, which by means of a least-squares fit were determined such that all three curves collapse. 
	The black lines in the lower panel represent the mean field (MF) approximations \eq{thetaMF} and \eq{thetaMF2}, valid for $X=(\mu-\mu_\text{c})/(mgT)\gg1$ and $X\ll-1$, respectively.  
	The dashed lines show the MF behavior for $\zeta_{\rho}=380$ and $\zeta_{\mu}=13.2$ of \cite{Prokofev2001a.PRL87.270402}, the solid lines for the values \eq{zetarho} and \eq{zetamu}.
	}
	\label{fig:eos}
\end{figure}

As was outlined in \Sect{BKT}, the weakly interacting Bose gas is expected to feature universal behavior in a wide range of the phase diagram, across the BKT transition and also further away from it. 
In order to analyze our data with respect to such universality, we depict, in the upper panel of \Fig{eos}, the equation of state $\rho(\mu)$ for a set of chemical potentials as obtained from our simulations for three different coupling strengths $mg\in\{0.025,0.1,0.2\}$, with $\mu$ given in units of $mgT$.
The data points are spline-interpolated for better visibility.
The respective critical $\mu_\mathrm{c}$ are indicated as thin vertical lines above the horizontal axis.
Since most data points are sufficiently far away from criticality, also for $mg=0.2$ runaway trajectories were mostly absent. However, for the two data points closest to the transition point, they started to occasionally occur, in which case we excluded the respective run, whereby we could still obtain a reasonable convergence of the total particle number. 
The inset shows the unscaled data as functions of $\mu/T$.

As is demonstrated in the lower panel of \Fig{eos}, shifting the curves by the respective $\rho_\text{c}\lambda_{T}^{2}$ downwards and $\mu_\text{c}/mgT$ to the left, with the $\rho_\text{c}$ and $\mu_\text{c}$ obtained as specified below, all curves collapse, according to \eq{eosuniversal} to a single scaling function $F$.
Hence our data corroborates the predicted scaling within the depicted window around the critical point.
As similarly found for the classical simulations in \cite{Prokofev2002a.PRA66.043608}, the mean-field (MF) prediction \eq{EOS_mean_field}, \eq{thetaMF}  (dashed line) agrees remarkably well with our CL results below the BKT transition, $X=(\mu-\mu_\mathrm{c})/(mgT)>0$, when choosing the non-universal constant $\zeta_{\mu}=13.2$ of Ref.~\cite{Prokofev2001a.PRL87.270402}.
As in the classical simulations, this is the case both, outside the fluctuation region, i.e., for $|X|\gtrsim1$, and in the close vicinity of the transition.
In contrast to this, above the transition, $X<0$, our results deviate from the mean-field prediction \eq{thetaMF2} for the same value of $\zeta_{\mu}$ (dashed line).

Since the definitions \eq{critical_dense} and \eq{critical_mu} of the constants $\zeta_{\rho}$ and $\zeta_{\mu}$, respectively, rest, however, on mean-field expressions for the scales $k_\text{c}$ and $k_{T}$, for our data to be consistent with the MF prediction, \eq{thetaMF}, \eq{thetaMF2}, outside the fluctuation region rather requires the different value \eq{zetamu} of the non-universal constant $\zeta_{\mu}$, cf.~the solid lines in the lower panel of \Fig{eos}.
Recall furthermore that the function $F$, according to \Eq{EOS_mean_field}, depends on the quotient $\zeta_{\rho}/\zeta_{\mu}$, which is understood to be independent of the UV details of the quantum gas and determines the value of the constant $\theta_{0}=\theta(0)=\ln(\zeta_{\rho}/\zeta_{\mu})/\pi$ \cite{Prokofev2002a.PRA66.043608}.
Hence, since our result for the dependence of $\rho_\mathrm{c}$ on $mg$ gave the value \eq{zetarho}, also a modified value of $\zeta_{\mu}$ is required for $F$ to remain unchanged outside the fluctuation region and below the transition, $X\gtrsim1$, as suggested by our results coinciding, in this regime, with the findings of \cite{Prokofev2002a.PRA66.043608}.

We emphasise that our data is obtained in the weakly interacting regime, $0.025\leq mg\leq0.2$, where the MF expressions \eq{thetaMF}, \eq{thetaMF2}, which are also based on the approximation that $mg\theta\ll1$ \cite{Prokofev2002a.PRA66.043608}, should be valid.  
As a result, our data yields a value of the universal constant
\begin{align}
	\theta_{0}=\pi^{-1}\ln(\frac{\zeta_{\rho}}{\zeta_{\mu}})
	&=0.97 \pm 0.01
	\label{eq:theta0}
	\,,
\end{align}
as compared with $\theta_{0}=1.068\pm0.01$ reported in \cite{Prokofev2002a.PRA66.043608}.
In summary, notwithstanding the modification of both, $\zeta_{\rho}$ and $\zeta_{\mu}$ as compared with the classical result, the universal function determining the equation of state remains unchanged within errors below the BKT transition, while it is slightly increased above, where quantum fluctuations are thus found to be more significant.

Under the approximation that the universal scaling form \eq{eosuniversal} holds, at least for the not too large values $mg$ considered here, we can also determine the shift in the critical chemical potentials and densities between different $mg$, by fitting the curves to collapse onto each other.
This allows extracting the critical chemical potentials and densities independent of the procedure described in the previous section once $\mu_\mathrm{c}$ and $\rho_\mathrm{c}$ are known for one value of $mg$. 
The $\mu_\mathrm{c}$ and $\rho_\mathrm{c}$ employed in this section were obtained in this way, with $\mu_\mathrm{c}(mg=0.1)=(1.531\pm0.006)\,mgT$ and $\rho_\mathrm{c}(mg=0.1)=(7.874\pm0.008)\lambda_T^{-2}$ taken from the $256^2$ lattice results of \Sect{critdens}. 
For a comparison of the critical densities obtained in this way, with those from \Sect{critdens}, see \App{critdens}.

\subsection{\label{sec:corrfunc}Single-particle momentum spectra}
%
\begin{figure}
	\includegraphics[width=0.9\columnwidth]{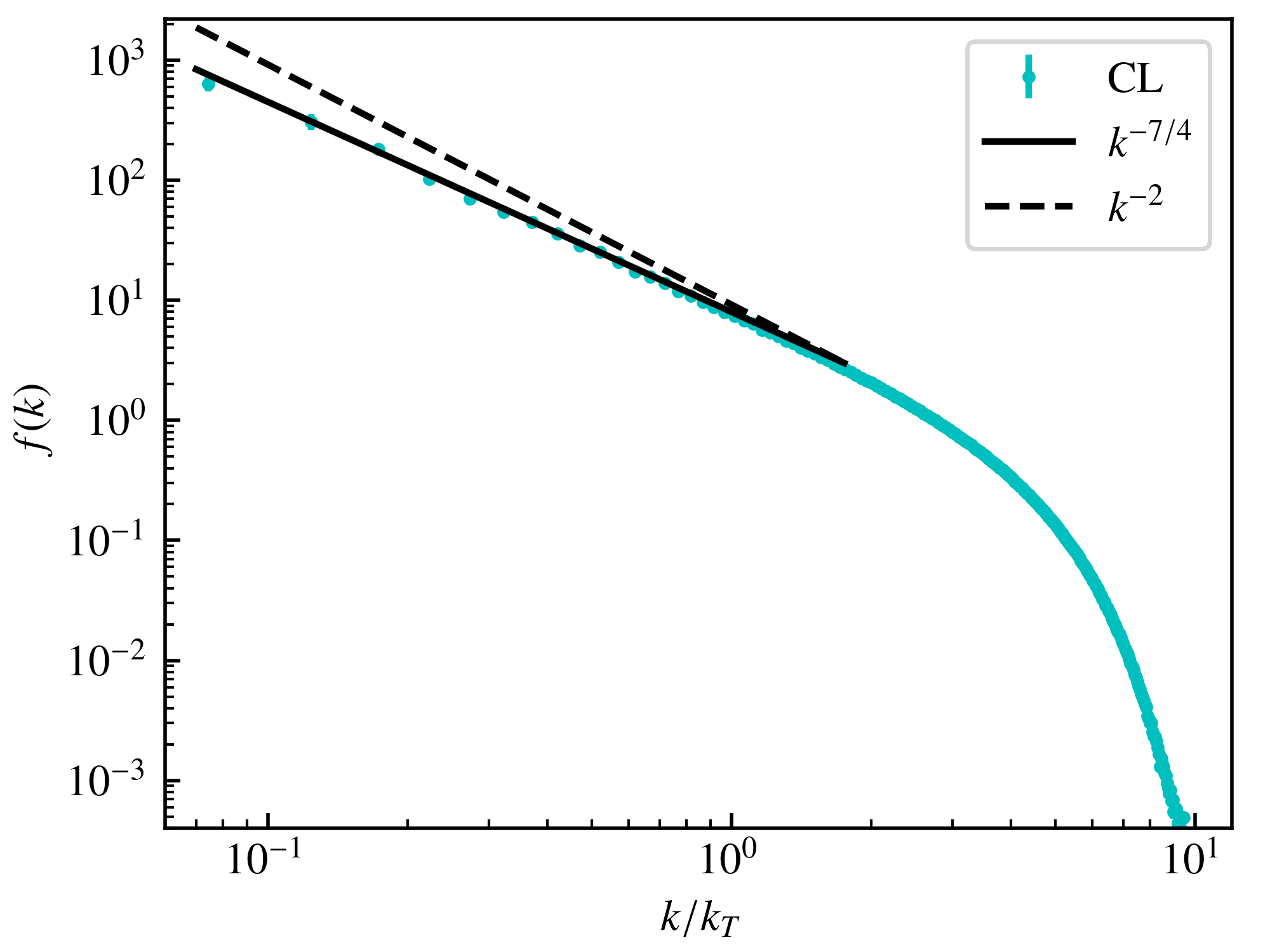}
	\includegraphics[width=0.9\columnwidth]{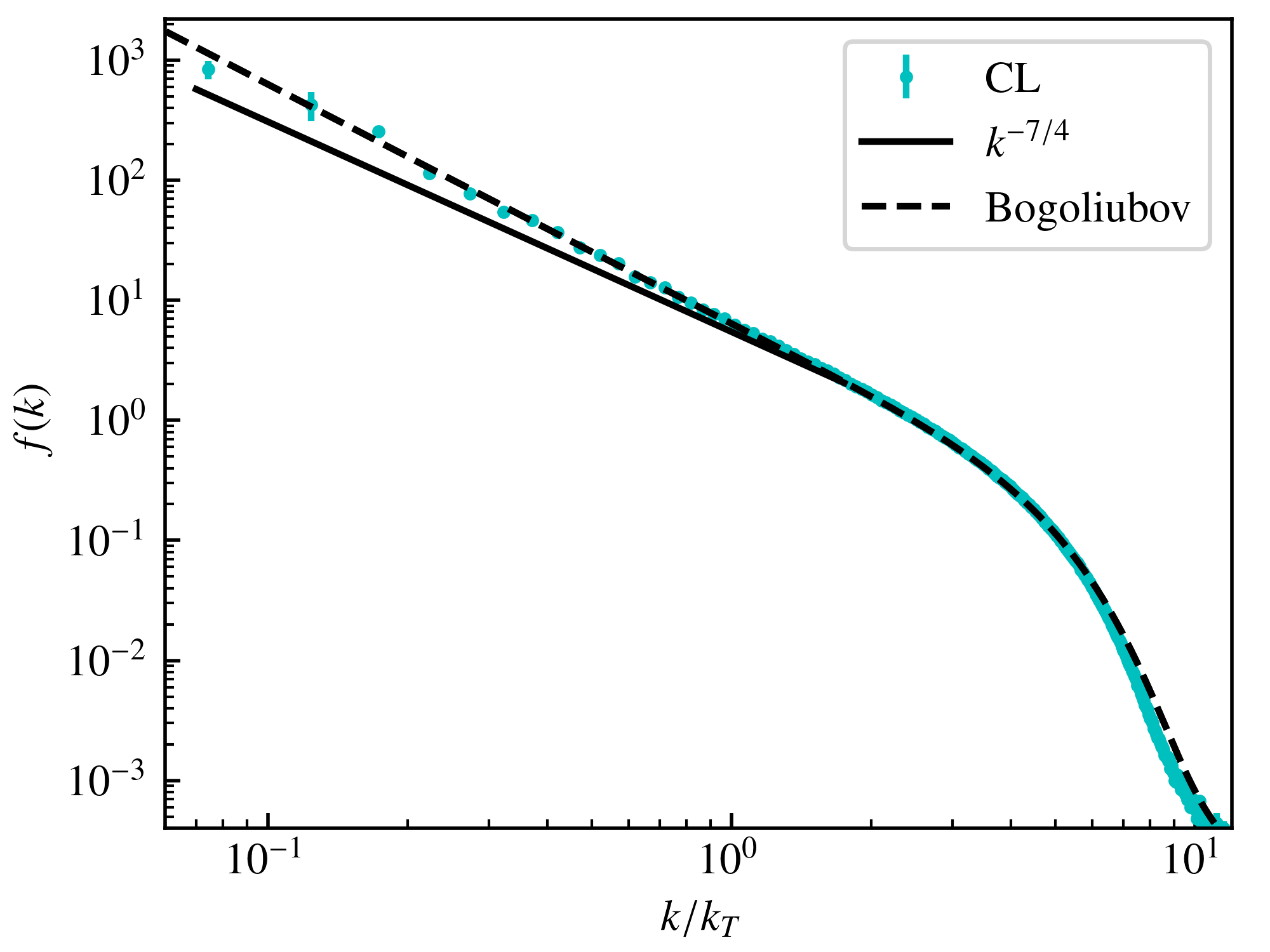}
	\caption{	
	Single-particle momentum spectra $f({k})=f(|\mathbf{k}|)$, \Eq{fk}, for a coupling $mg=0.1$ and two different chemical potentials: $\mu/(mgT)=2$, slightly below the transition (upper panel), and $\mu/(mgT)=4$, far below the transition (lower panel). 
	The momentum is expressed in units of the thermal momentum $k_T=1/\lambda_T$. 
	Whereas slightly below the transition the spectrum approximately shows a $k^{-7/4}$ power law in the infrared, below the thermal wave number $k_{T}$, consistent with the near-critical scaling predicted by BKT theory, we find a mean-field Rayleigh-Jeans $\sim k^{-2}$ fall-off further below the transition, which forms part of the Bogoliubov distribution (dashed line). The condensate fractions $f(\mathbf{k}=0)/N_\text{tot}$ are $61\%$ and $82\%$, respectively, i.e. due to the finite extent of our system we find a macroscopic occupation of the condensate mode.
	}
	\label{fig:spectrum}
\end{figure}
\begin{figure}
	\includegraphics[width=0.9\columnwidth]{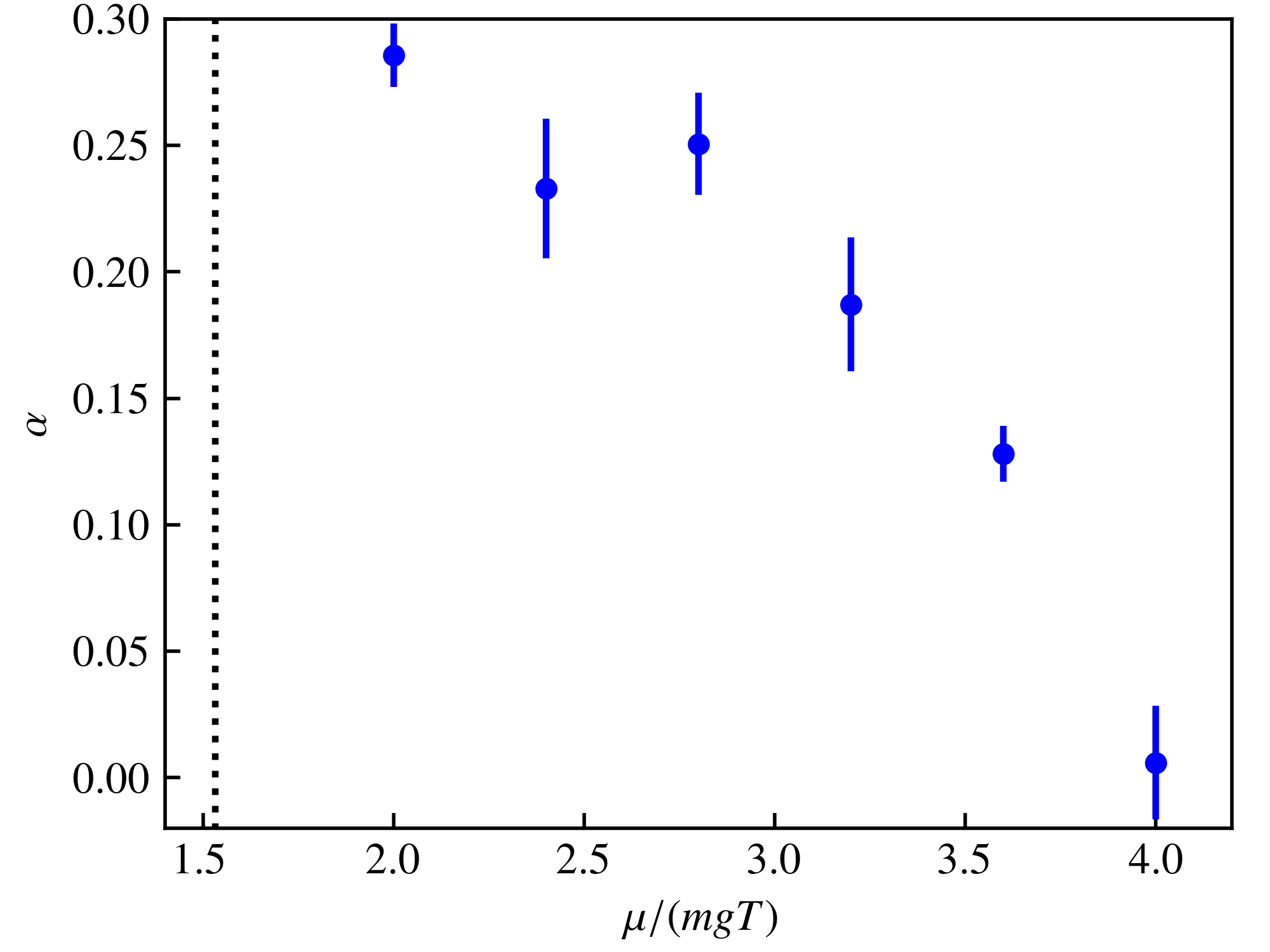}
	\caption{ 
	The power-law exponent $\alpha$ as a function of $\mu/(mgT)$, as obtained from a least-squares fit of a linear function to $\log f(k)$ vs. $\log k$, obtained from our simulations for $mg=0.1$. The critical chemical potential is marked by a dotted black line. Going from close to the transition to far below the transition, $\alpha$ decreases from $\simeq1/4$ to zero, in accordance with the prediction from BKT theory.
	}
	\label{fig:alphas}
\end{figure}

In this section, we use the CL data to evaluate the single-particle momentum-dependent occupation number distribution \eq{fk} for comparison with the infrared scaling \eq{fkscaling} expected close to and further below the BKT transition. 
We performed simulations for a coupling $mg=0.1$ at several chemical potentials ranging from $\mu/(mgT)=2$ to $\mu/(mgT)=4$, i.e., from slightly below to far below the BKT transition. 
The results for $\mu/(mgT)=2$ and $4$ are shown in \Fig{spectrum}. 

As predicted by BKT theory, far below the transition, the distribution exhibits a $\sim k^{-2}$,  Rayleigh-Jeans power law in the infrared regime of momenta far below the temperature scale, while it falls off exponentially at larger momenta.
This is consistent with a thermal distribution of Bogoliubov quasiparticles, resulting in the dashed distribution in the lower panel.
Close to the BKT transition, in contrast, the power law in the infrared is slightly reduced to $\sim k^{-7/4}$. 
In order to make this more systematic, we have extracted $\alpha$ for a range of chemical potentials by fitting a linear function to $\log f(k)$ vs.~$\log k$ in the power-law region. 
The results are shown in \Fig{alphas}. 
As one can see, the value of $\alpha$ drops  from $\alpha\simeq1/4$ towards zero, between the chemical potentials slightly below and far below the transition, in accordance with the BKT prediction.

\subsection{\label{sec:vortices}Vortex unbinding across the BKT transition}
%
\begin{figure}
	\includegraphics[width=0.9\columnwidth]{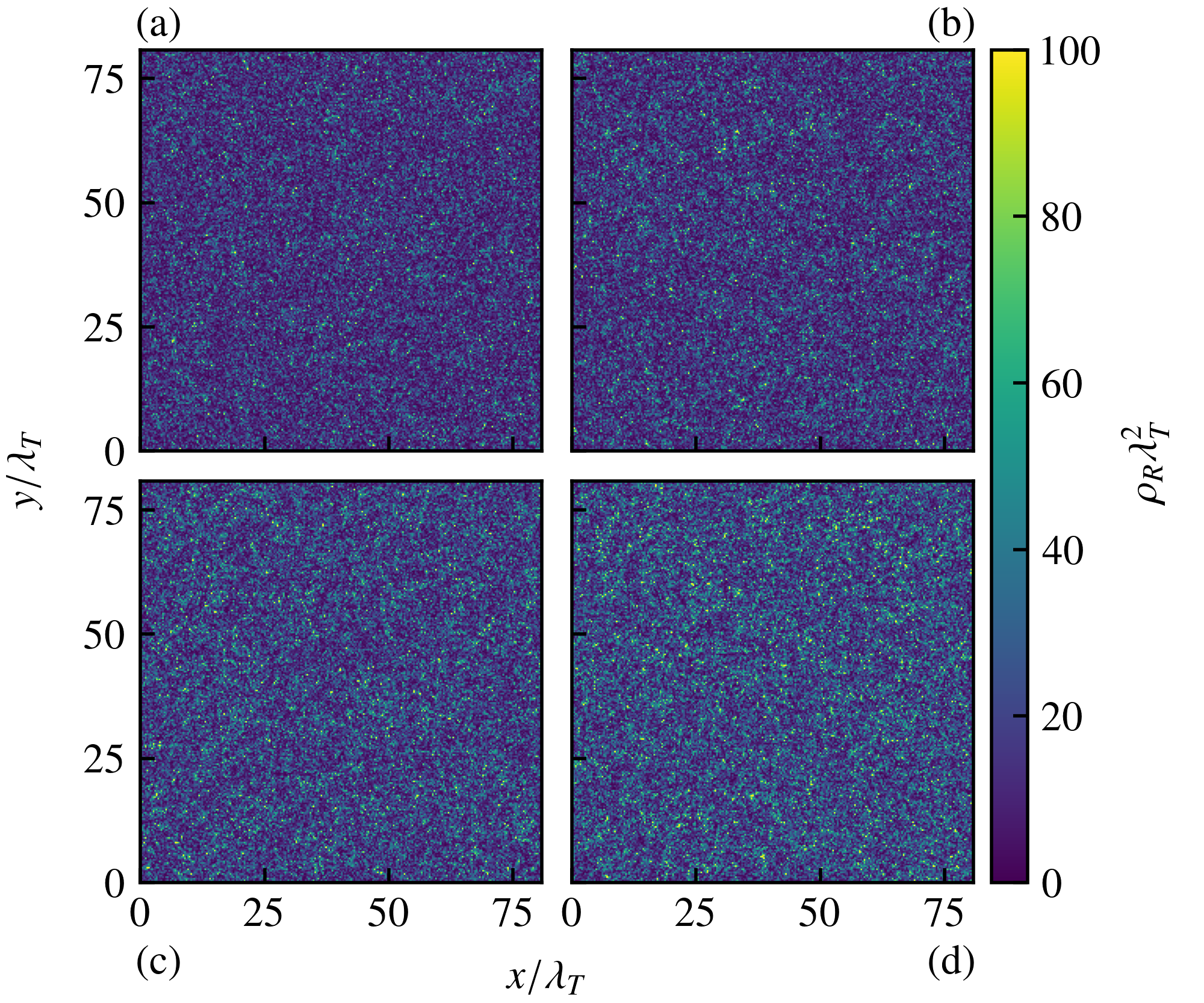}
	\includegraphics[width=0.9\columnwidth]{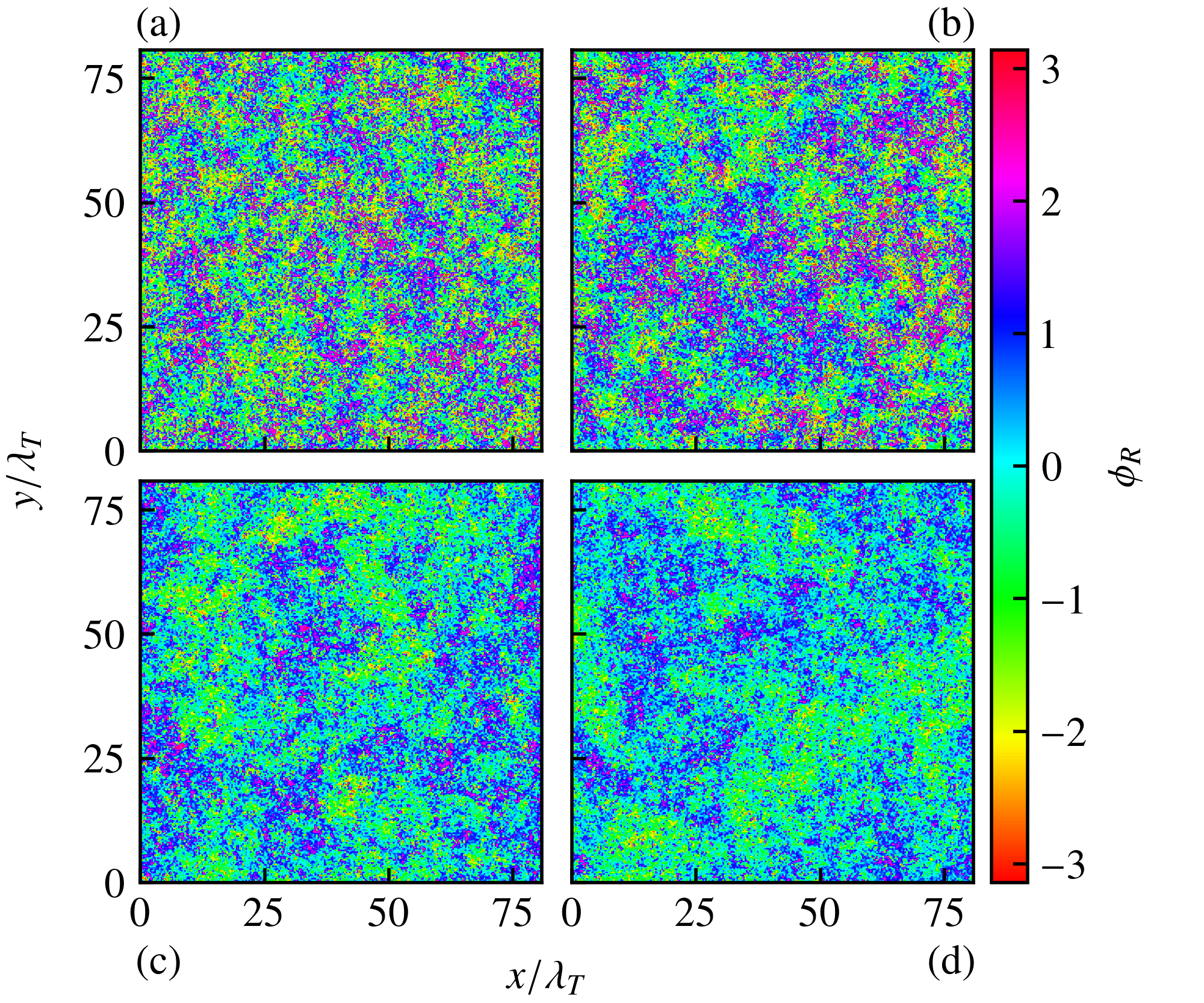}
	\caption{ 
		Snapshots of the position-space density $\rho_R\equiv \varphi_R^2+\chi_R^2$ at $\vartheta=4\cdot 10^3 \,a_\text{s}^{-3}$ for $mg=0.1$ and four different chemical potentials (a) $\mu/(mgT)=0.8$, (b) $1.2$, (c) $1.6$, and (d) $2.0$ (upper panel). 
		The temperature is kept fixed while the chemical potential is varied, such that the healing length $\xi_\text{h}\equiv1/\sqrt{2m\mu}$ in units of the lattice spacing varies between $\xi_\text{h}=3.16\,a_\text{s}$ and $\xi_\text{h}=2\,a_\text{s}$. 
		The mean field densities $\lambda_T^2 \bar{\rho}\equiv \lambda_T^2 \mu/g$ are $\lambda_T^2 \bar{\rho}=5.02,7.54,10.05,12.56$. 
		As one can see, the position-space densities are completely dominated by fluctuations such that it is difficult to infer information about the topological phase transition from them. 
		The lower panel shows the position-space phase $\phi_R\equiv \arg\left(\varphi_R+\mathrm{i}\chi_R\right)$ for the same parameters, which indicates phase ordering across the transition.
	}
	\label{fig:realspace}
\end{figure}
\begin{figure}
	\includegraphics[width=0.9\columnwidth]{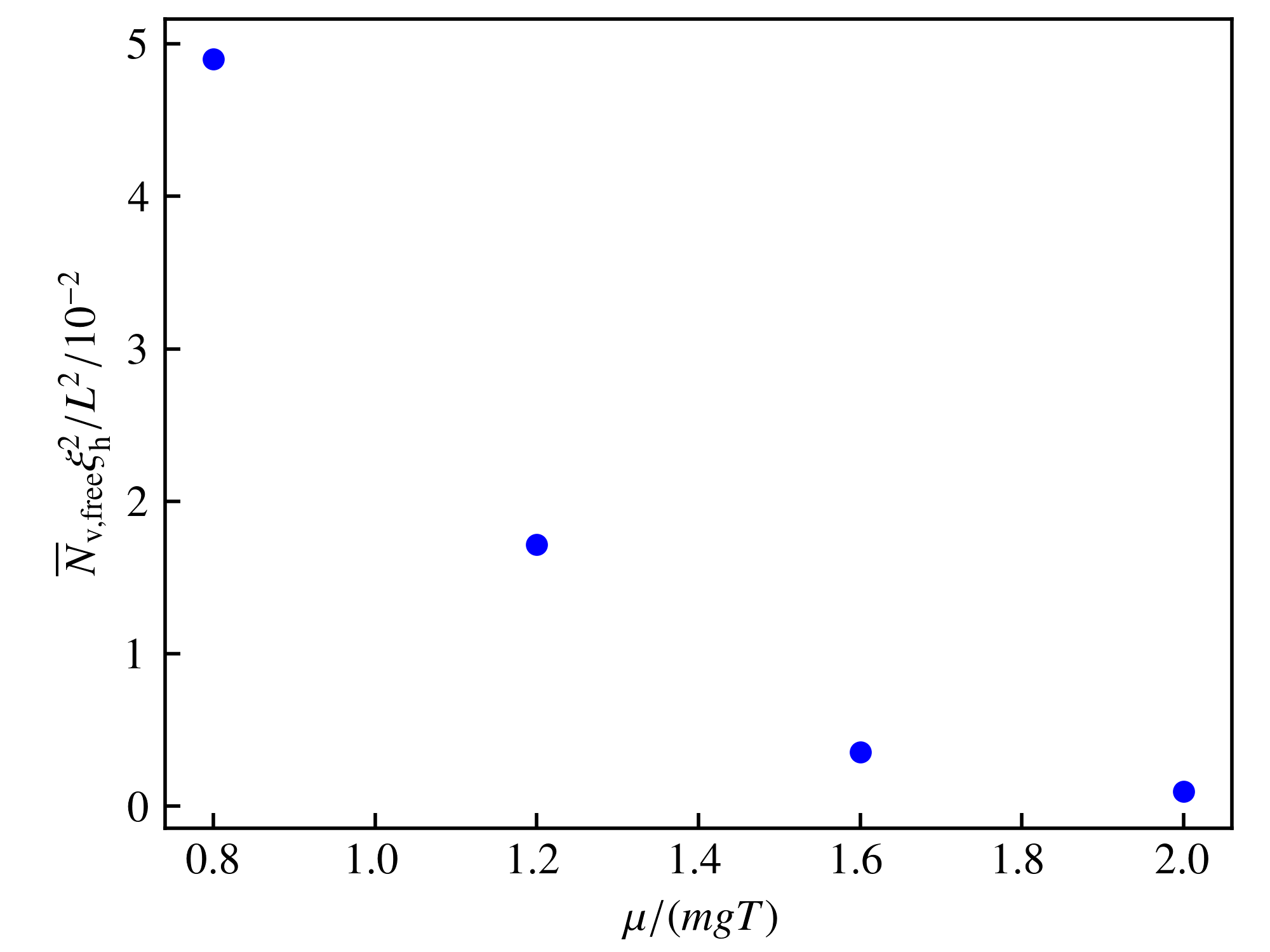}

	\caption{ 
		 The mean integrated rotational part of the current squared, $\rho_{\mathrm{v,free}}\xi_\mathrm{h}^{2}$, as defined in \eq{defY}, as a function of $\mu/(mgT)$, for $mg=0.1$. 
		 It can be considered as a measure for the mean number of unbound vortices appearing in the system, see the discussion in the main text. Error bars are too small to be visible.
	}
	\label{fig:jrot2}
\end{figure}
One of the key features of the BKT transition is that it is characterized by the transition from a free vortex gas to bound vortex-antivortex pairs, which we here attempt to study within the CL framework. 
Let us first have a look at the configurations produced by the Langevin process in position space. 
We consider the density $\rho_R$ and phase $\phi_R$ characterising the real parts of the complexified two-component field, defined as 
\begin{align}
	\rho_R&\equiv\varphi_R^2+\chi_R^2
	\,,&
	\phi_R&\equiv\arg(\varphi_R+i\chi_R)
	\,.
\end{align}
Their distribution across the spatial grid are shown in \Fig{realspace}, for four values of the chemical potential $\mu$ in the vicinity of the BKT transition. 
While the density $\rho_R$ seems entirely dominated by noise, the phase $\phi_R$ reveals the phase ordering process across the transition.
However, one does not observe stable vortices in the position-space configurations. 
This is not a genuine property of complex Langevin simulations, but is also the case e.g. for ordinary Monte Carlo simulations of the XY model, cf.~Fig.~1 in \cite{Peled2019a.Sojourns}.

\begin{figure}
	\includegraphics[width=0.9\columnwidth]{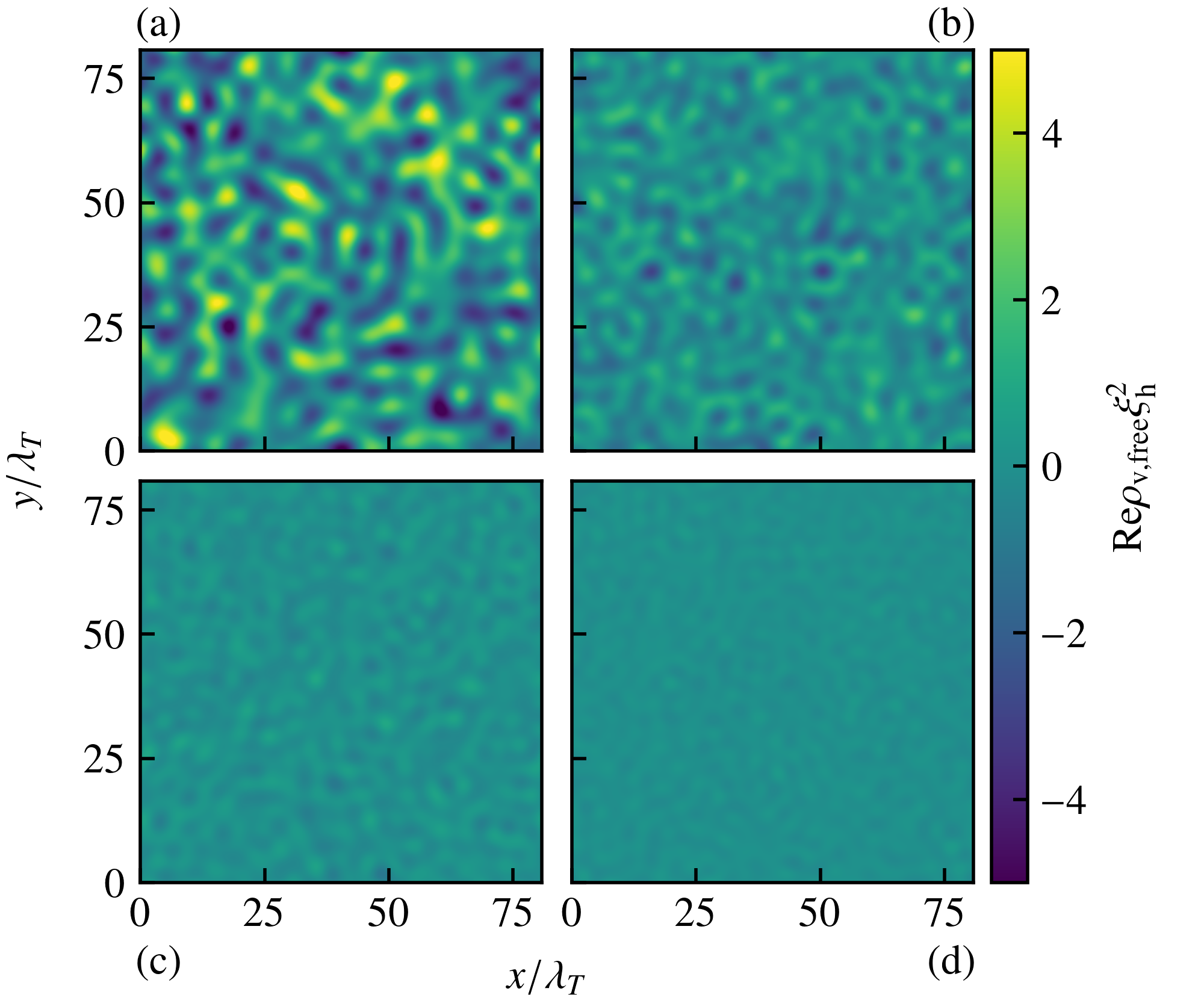}
	\caption{ 
		Snapshots of the real part of $\rho_{\mathrm{v,free}}(\mathbf{x})\xi_\mathrm{h}^2$, as defined in \eq{defrhovfree} and \App{evY} for $mg=0.1$ and the same four different chemical potentials (a) $\mu/(mgT)=0.8$, (b) $1.2$, (c) $1.6$, and (d) $2.0$, corresponding to the data shown in \Fig{realspace}. For a better visibility, fluctuations above the healing momentum $k_\mathrm{h}=\sqrt{2m\mu}$ are filtered out by a low-pass filter. Note that by virtue of the complexification prescription of the CL algorithm, $\rho_{\mathrm{v,free}}(\mathbf{x})\xi_\mathrm{h}^2$ will be in general complex in single realizations and its real part can become negative, while in the long-time average, $\rho_{\mathrm{v,free}}(\mathbf{x})\xi_\mathrm{h}^2$ will come out real and positive.
	}
	\label{fig:Ytilde}
\end{figure}

One may conclude that expectation values, i.e.~long-time averages along the Langevin trajectories, must be considered in order to gain insight into the topological phase transition. 
Here, one encounters a CL-specific difficulty. 
The analysis of topological properties of a Bose gas usually requires determining the phase of the complex field $\psi$ or the velocity field $\mathbf{v}=\mathbf{j}/\rho$. 
Both lead to observables that are non-holomorphic in the fields. 
The CL algorithm, however, requires the real-valued fields to be analytically continued, such that the action and observables should preferably be holomorphic. Non-holomorphic actions and observables lead to both computational problems (there can be numerical divergencies when the trajectories come close to the singularities) as well as conceptual ones (the correctness of the method is more difficult to establish) \cite{Aarts2017a.JHEP5.1.complex}.
As a result, no vortex positions and numbers can be determined from the data depicted in \Fig{realspace}.
This clearly represents a limitation to the applicability of the CL algorithm in this case. 
For this reason we resort to evaluating the current density instead of the velocity field,  as discussed in \Sect{TopoBKT}, which is perfectly holomorphic in $\varphi$ and $\chi$ but has the drawback that it is affected by fluctuations of the density.
In \Fig{jrot2}, we show the average
\begin{align}
	\label{eq:defY}
	\frac{\xi_\mathrm{h}^{2}}{L^2}\overline{N}_\text{v,free}
	&= \frac{\xi_\mathrm{h}^{2}}{L^2}\int \mathrm{d}^2x\,\left\langle\,\rho_{\mathrm{v,free}}(\mathbf{x})\right\rangle
	\,,
\end{align}
with healing length $\xi_\mathrm{h}=(2m\mu)^{-1/2}$, i.e., the mean number of free vortices per healing length squared, which can be considered to be a measure for the transition from a vortex gas to bound vortex-anti-vortex pairs, and with $\rho_{\mathrm{v,free}}(\mathbf{x})$ determined according to \eq{defrhovfree}.
Snapshots of spatial distribution of $\rho_{\mathrm{v,free}}(\mathbf{x})$ are depicted in \Fig{Ytilde}, for $mg=0.1$ and the same four chemical potentials as chosen in \Fig{realspace}. 
Details on the numerical evaluation of $\rho_{\mathrm{v,free}}$ can be found in \App{evY}.
The decrease of $\rho_{\mathrm{v,free}}$ across the transition corroborates the vortex-anti-vortex recombination process, when going over from the disordered above to the ordered phase below the transition.

\section{Summary and outlook}
Employing the complex Langevin algorithm, we have performed an ab-initio simulation of the interacting Bose gas in two spatial dimensions across the BKT phase transition. 
We have found that the CL method is able to successfully reproduce central characteristics of BKT physics, namely the universality in the equation of state, the algebraic decay of correlation functions and the vortex-unbinding mechanism. 
We have furthermore analysed the dependence of the critical density on the interaction strength $mg$ and compared to  results from simulations of the classical field theory \cite{Prokofev2001a.PRL87.270402}. 
Our simulations of the full quantum model yield small but significant deviations from the classical-field-theory predictions.

Hence, this work demonstrates that the CL algorithm can be a viable tool for performing full quantum simulations of the topological phase transition in a weakly interacting Bose gas, i.e., could be of use when quantum corrections beyond classical Gross-Pitaevskii simulations play a role. 
Experimentally relevant applications include the BKT transition in external trapping potentials \cite{Hadzibabic2006a,Bisset2009a.PRA79.033626}, with long-range interactions \cite{Fedichev2012a.PPNL9.71,Giachetti2022a.PRB106.014106} or in multi-component gases \cite{Kasamatsu2005a.IJMPB19.1835,Kobayashi2019a.JPSJ88.094001,Furutani2023PhysRevA.107.L041302}.

\begin{acknowledgments}
The authors thank Felipe Attanasio, Karthik Chandrashekara, Lauriane Chomaz, Martin G\"arttner, Nikolas Liebster, Markus Oberthaler, Jan Pawlowski, Nikolay Prokof'ev, Adam Rançon, Alessio Recati, Santo-Maria Roccuzzo, Marius Sparn, Helmut Strobel, and Boris Svistunov for discussions and collaboration on related topics. 
This work is supported by the German Research Foundation (Deutsche Forschungsgemeinschaft, DFG) under Germany's Excellence Strategy EXC 2181/1- 390900948 (the Heidelberg STRUCTURES Excellence Cluster), under SFB 1225 ISOQUANT - 273811115, as well as grant GA677/10-1.
The authors furthermore acknowledge support by the state of Baden-W{\"u}rttemberg through bwHPC and DFG through
grants INST 35/1134-1 FUGG, INST 35/1503-1 FUGG, INST 35/1597-1 FUGG, and 40/575-1 FUGG.
\end{acknowledgments}

\begin{appendix}
\begin{center}
\textbf{APPENDIX}
\end{center}
%
%
\section{Discretized CL equations  \label{app:CLequations}} 
In this appendix, we provide details of the discretized CL equations of motions, Eqs.~\eq{cl_eq}, \eq{cl_eq_Im}, for the case of a Bose gas described by the action \eq{action} in two spatial dimensions.
See \Cite{Heinen2022a.PhysRevA106.063308.complex} for a discretized version of the (3D) action.
We express the complex Bose fields in terms of its real and imaginary parts as 
\begin{align}
	\psi\equiv\varphi+\mathrm{i}\chi
	\,. 
\end{align}
Discretizing these components on the $N_\mathrm{s}^{2}\times N_{\tau}$ lattice, $\varphi(\tau,\mathbf{x})=\varphi(ia_{\tau},\mathbf{n}a_\mathrm{s})\equiv\varphi_{i,\mathbf{n}}$, etc. for $\chi$, where the index $i$ enumerates the imaginary-time lattice sites with spacing $a_\tau$, and the three-dimensional index vector $\mathbf{n}$  the spatial lattice sites, with discretization $a_\mathrm{s}$. 
The CL are obtained from the discretized action \cite{Heinen2022a.PhysRevA106.063308.complex} by taking derivatives with respect to $\varphi$ and $\chi$, 
\begin{align}
	\label{eq:deriv1}
	\frac{\partial\varphi_{i,\mathbf{n}}}{\partial \vartheta}
	=&\ a_\mathrm{s}^2\,\left(
	\varphi_{i+1,\mathbf{n}}
	+\varphi_{i-1,\mathbf{n}}
	-2\varphi_{i,\mathbf{n}}
	+\mathrm{i}\chi_{i-1,\mathbf{n}}
	-\mathrm{i}\chi_{i+1,\mathbf{n}}
	\right)
	\nonumber\\
	&+\frac{a_\tau}{2m}\,\Delta^\text{lat}\bigg(
	\varphi_{i-1,\mathbf{n}}
	+\varphi_{i+1,\mathbf{n}}
	+\mathrm{i}\chi_{i-1,\mathbf{n}}
	-\mathrm{i}\chi_{i+1,\mathbf{n}}
	\bigg)
	\nonumber\\
	&+\mu a_\mathrm{s}^2a_\tau\,\bigg(
	\varphi_{i-1,\mathbf{n}}
	+\varphi_{i+1,\mathbf{n}}
	+\mathrm{i}\chi_{i-1,\mathbf{n}}
	-\mathrm{i}\chi_{i+1,\mathbf{n}}
	\bigg)
	\nonumber\\
	&-ga_\mathrm{s}^2a_\tau\,\bigg(
	\Psi_{i-1,\mathbf{n}}\,\varphi_{i-1,\mathbf{n}}
	+\Psi_{i,\mathbf{n}}\,\varphi_{i+1,\mathbf{n}}
	\nonumber\\
	&\qquad 
	+\mathrm{i}\Psi_{i-1,\mathbf{n}}\,\chi_{i-1,\mathbf{n}}
	-\mathrm{i}\Psi_{i,\mathbf{n}}\,\chi_{i+1,\mathbf{n}}
	\bigg)+\eta(\vartheta)
	\,,\\
%
%
	\label{eq:deriv2}
	\nonumber
	\frac{\partial\chi_{i,\mathbf{n}}}{\partial \vartheta}
	=&\ a_\mathrm{s}^2\,\left(
	\chi_{i+1,\mathbf{n}}
	+\chi_{i-1,\mathbf{n}}
	-2\chi_{i,\mathbf{n}}
	-\mathrm{i}\varphi_{i-1,\mathbf{n}}
	+\mathrm{i}\varphi_{i+1,\mathbf{n}}
	\right)
	\nonumber\\
	&+\frac{a_\tau}{2m}\,\Delta^\text{lat}\bigg(
	\chi_{i-1,\mathbf{n}}
	+\chi_{i+1,\mathbf{n}}
	-\mathrm{i}\varphi_{i-1,\mathbf{n}}
	+\mathrm{i}\varphi_{i+1,\mathbf{n}}\bigg)
	\nonumber\\
	&+\mu a_\mathrm{s}^2a_\tau\,\bigg(
	\chi_{i-1,\mathbf{n}}
	+\chi_{i+1,\mathbf{n}}
	-\mathrm{i}\varphi_{i-1,\mathbf{n}}
	+\mathrm{i}\varphi_{i+1,\mathbf{n}}\bigg)
	\nonumber\\
	&-ga_\mathrm{s}^2a_\tau\,\bigg(
	\Psi_{i-1,\mathbf{n}}\,\chi_{i-1,\mathbf{n}}
	+\Psi_{i,\mathbf{n}}\,\chi_{i+1,\mathbf{n}}
	\nonumber\\
	&\qquad 
	-\mathrm{i}\Psi_{i-1,\mathbf{n}}\,\varphi_{i-1,\mathbf{n}}
	+\mathrm{i}\Psi_{i,\mathbf{n}}\,\varphi_{i+1,\mathbf{n}}
	\bigg)+\eta(\vartheta)\,,
\end{align}
where
\begin{align}
	\Psi_{i,\mathbf{n}}
	&\equiv
	\varphi_{i+1,\mathbf{n}}\,\varphi_{i,\mathbf{n}}
	+\chi_{i+1,\mathbf{n}}\,\chi_{i,\mathbf{n}}
	\nonumber\\
	&\qquad
	+\mathrm{i}\varphi_{i+1,\mathbf{n}}\,\chi_{i,\mathbf{n}}
	-\mathrm{i}\varphi_{i,\mathbf{n}}\,\chi_{i+1,\mathbf{n}}
	\,.
	\label{eq:Psi}
\end{align}
$\Delta^\text{lat}$ is the Laplacian on the lattice, for which there are several possible choices. 
One may approximate the Laplacian by finite differences, i.e., as
\begin{align}
	\Delta^\text{lat,FD}A_{i,\mathbf{n}}
	\equiv&\ 
	A_{i,\mathbf{n}+\mathbf{e}_x}+A_{i,\mathbf{n}-\mathbf{e}_x}+
	A_{i,\mathbf{n}+\mathbf{e}_y}+A_{i,\mathbf{n}-\mathbf{e}_y}
	-4A_{i,\mathbf{n}}
	\,,
	\label{eq:LaplaceLatt}
\end{align}
with $\mathbf{e}_{x,y}$ being the unit vectors in two dimensions.
Here we rather choose to evaluate the Laplacian via a spectral derivative, 
\begin{align}
	\Delta^\text{lat,sp}A_{i,\mathbf{n}}
	\equiv& -\frac{1}{N_\text{s}^2}\sum_{\mathbf{m}\mathbf{n}'}
	\exp\left(\mathrm{i}\frac{2\pi}{N_\text{s}}\mathbf{m}\cdot(\mathbf{n}-\mathbf{n}')\right)\,
	\left|\frac{2\pi\mathbf{m}}{N_\text{s}}\right|^2\,A_{i,\mathbf{n}'}
	\,.
	\label{eq:LaplaceLattSpect}
\end{align}
While the spectral derivative is numerically more expensive than finite differences because it involves two FFTs, it has the advantage that it reproduces the correct quadratic continuum dispersion, whereas for the finite difference discretization one obtains a sine-like lattice dispersion.
Thus, it is not necessary to take into account the sine-spacing of momenta in the evaluation of observables, as was done in \cite{Heinen2022a.PhysRevA106.063308.complex}, which simplifies the computation of quantities like the superfluid density and the currents discussed in the subsequent appendices.

Note furthermore that, in Eqs.~\eq{deriv1}--\eq{Psi}, $\psi^*$ must be evaluated infinitesimally later than $\psi$, i.e.~at lattice point $i+1$ rather than at $i$. This follows from the construction of the coherent-state path integral where, in each time step, $a^\dagger$ is evaluated with respect to the coherent state on the left while $a$ acts to the right, on the state one step earlier. 
This is important even in the analytical calculations where it ensures the right convergence in the complex Matsubara plane \cite{Altland2010a}.

\section{Evaluation of the superfluid density \label{app:supdens}} 
For the extraction of the superfluid density, we need the variance of the total momentum $\mathbf{P}$ which is measured by the operator
\begin{align}
\mathbf{P}=\sum_\mathbf{p} a_\mathbf{p}^\dagger a_\mathbf{p}\,\mathbf{p}\,.
\end{align}
The operator $\mathbf{P}^2$ can then be written as 
\begin{align}
	\nonumber\mathbf{P}^2
	&=\sum_{\mathbf{p}\mathbf{q}} a_\mathbf{p}^\dagger a_\mathbf{p} a_\mathbf{q}^\dagger a_\mathbf{q}\,\mathbf{p}
	\cdot\mathbf{q}
	\\
	&=\sum_{\mathbf{p}\mathbf{q}} a_\mathbf{p}^\dagger a_\mathbf{q}^\dagger a_\mathbf{p} a_\mathbf{q}\,\mathbf{p}
	\cdot\mathbf{q}+\sum_{\mathbf{p}}a_\mathbf{p}^\dagger a_\mathbf{p}\,|\mathbf{p}|^2
	\,.
\end{align}
In the path integral formalism this translates into
\begin{align}
	\label{eq:P2}
	\langle\mathbf{P}^2\rangle
	=\left\langle\frac{1}{N_\tau}\sum_i\left\{\left|\sum_\mathbf{p}\psi_{i+1,\mathbf{p}}^*\psi_{i,\mathbf{p}}\,\mathbf{p}\right|^2
	+\sum_\mathbf{p}\psi_{i+1,\mathbf{p}}^*\psi_{i,\mathbf{p}}\,|\mathbf{p}|^2\right\}\right\rangle
	\,,
\end{align}
where $i$ numbers the imaginary time slices and $N_\tau$ is their total number. 
Since we compute the Laplacian by a spectral derivative, \eq{LaplaceLattSpect}, the momenta can be defined as in the continuum, i.e. $\mathbf{p}=2\pi\mathbf{m}/N_\text{s}a_\text{s}$, with $\mathbf{m}$ the index of the corresponding mode.
Similarly, the total particle number is computed as 
\begin{align}
	\langle N\rangle
	=\left\langle\frac{1}{N_\tau}\sum_i\sum_\mathbf{p}\psi_{i+1,\mathbf{p}}^*\psi_{i,\mathbf{p}}\right\rangle
	\,.
\end{align}
%

\begin{figure}[h!]
	\includegraphics[width=0.9\columnwidth]{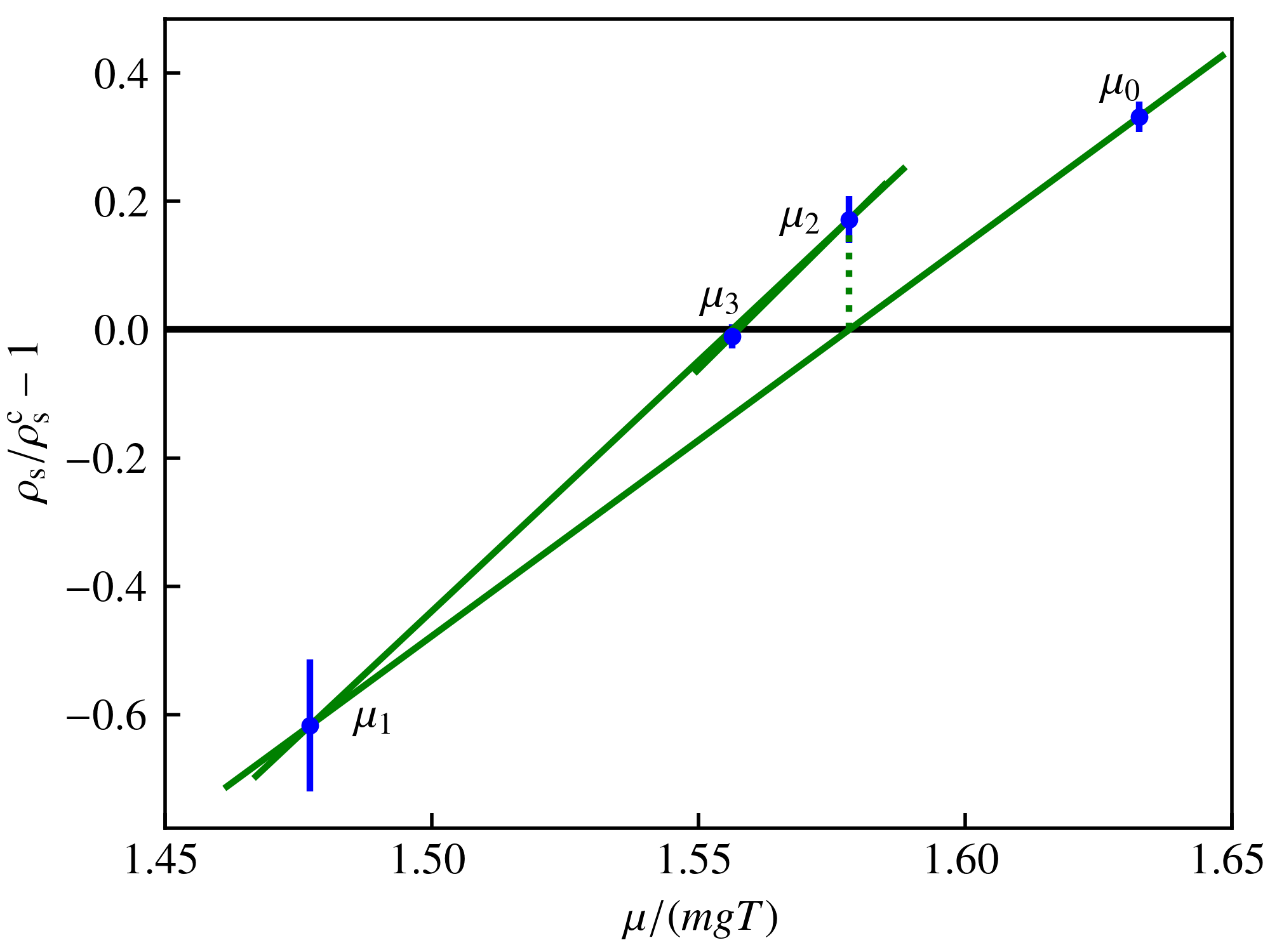}
	\includegraphics[width=0.9\columnwidth]{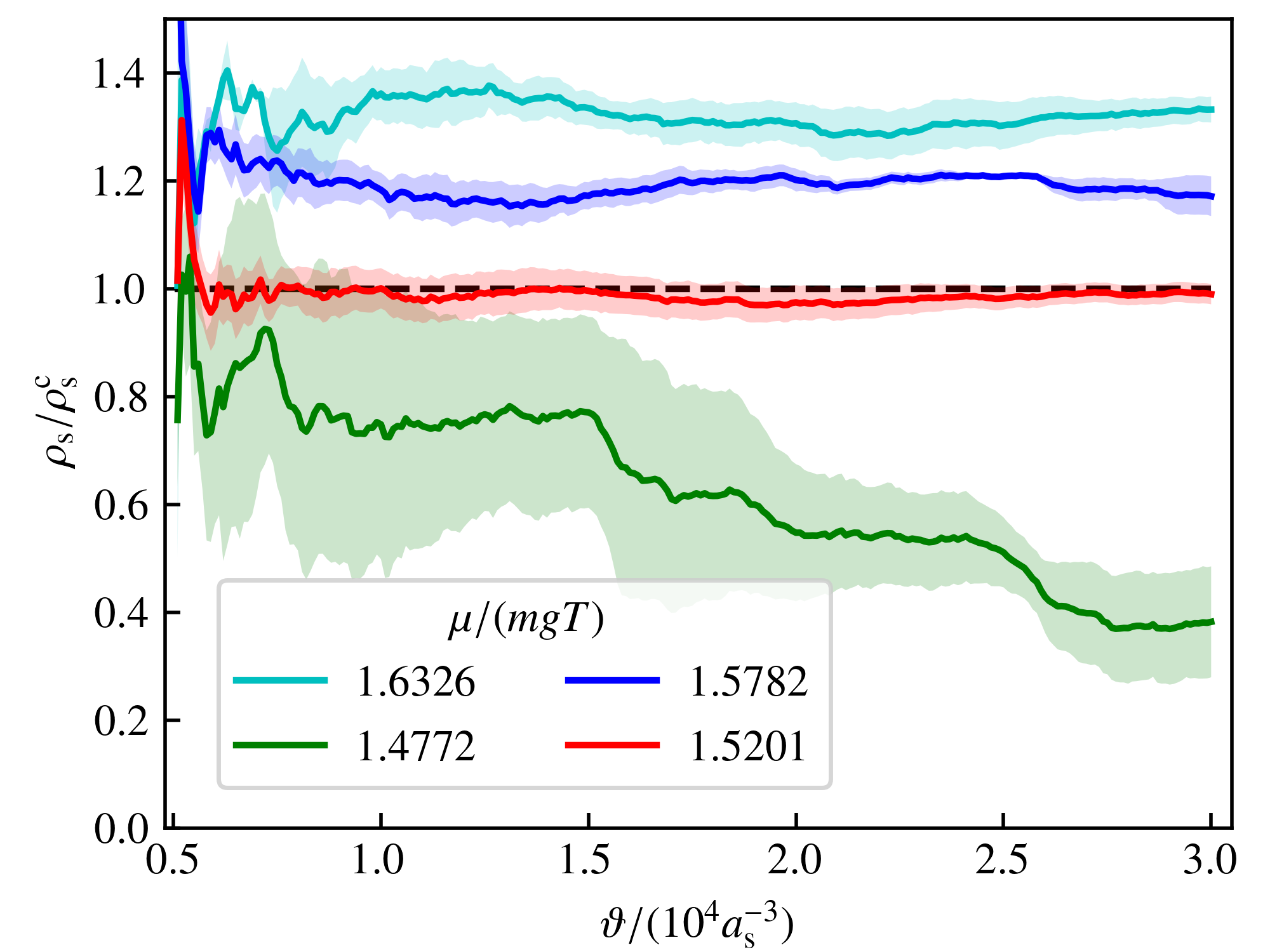}
	\caption{Determination of the BKT transition point by the secant method for $mg=0.1$ and $L/a_\mathrm{s}=128$. 
		Starting from two initial guesses for the critical chemical potential, $\mu_0/(mgT)=1.633$ $\mu_1/(mgT)=1.477$, subsequent guesses are chosen as the intersection of the secant through the previous two data points and the $\mu/T$-axis (upper panel). 
		The convergence is extremely fast. 
		The lower panel shows the convergence of the  superfluid densities as in Langevin time $\vartheta$, with error bands obtained from the variance of several statistically independent runs. 
	}	
	\label{fig:secant}
\end{figure}
%
%
\section{\label{app:evY}Evaluation of the vortex density}
In this appendix we provide details on the numerical extraction of the mean vortex density \eq{defrhovfree} and of the integrated vortex number \eq{defY} from the CL data. 
This is done most conveniently in momentum space, since both, the gradient appearing in the current, as well as the Helmholtz decomposition becomes a simple vector operation in momentum space. 
In momentum space, the current ${\mathbf{j}}_{i,\mathbf{p}}$, reads
\begin{align}
	\mathbf{j}_{i,\mathbf{p}}
	=\frac{1}{2m}\sum_\mathbf{q}(\mathbf{p}+2\mathbf{q})\,\psi_{i+1,\mathbf{p}+\mathbf{q}}^*\psi_{i,\mathbf{q}}
	\,
\end{align}
The Helmholtz decomposition, in momentum space, reads 
\begin{align}
	\mathbf{j}_{i,\mathbf{p}}
	=\mathbf{j}_{i,\mathbf{p}}^\text{irr}+\mathbf{j}_{i,\mathbf{p}}^\text{rot}
	=\mathbf{p}\frac{\mathbf{p}\cdot\mathbf{j}_{i,\mathbf{p}}}{|\mathbf{p}|^2}
	-\frac{\mathbf{p}\times\mathbf{p}\times\mathbf{j}_{i,\mathbf{p}}}{|\mathbf{p}|^2}
	\,.
\end{align}
The integral over the squared current density can be straightforwardly computed in momentum space as 
\begin{align}
	\int \mathrm{d}^2x\,\left\langle\,|\mathbf{j}_\text{rot}(\mathbf{x})|^2\right\rangle 
	&=\left\langle\frac{1}{N_\tau}\sum_i\sum_\mathbf{p}
	\,\mathbf{j}_{i,\mathbf{p}}^\text{rot}\cdot \mathbf{j}_{i,-\mathbf{p}}^\text{rot}\right\rangle
	\,.
\end{align}
For the position-space snapshots of $|\mathbf{j}_\text{rot}(\mathbf{x})|^2$ we Fourier transform $\mathbf{j}_{i,\mathbf{p}}^\text{rot}$, i.e. $\mathbf{j}_{i,\mathbf{x}}^\text{rot}=(N_\mathrm{s}a_\mathrm{s})^{-1}\sum_\mathbf{p} e^{\mathrm{i}\mathbf{x}\cdot\mathbf{p}}\mathbf{j}_{i,\mathbf{p}}^\text{rot}$ and define $|\mathbf{j}_\text{rot}(\mathbf{x})|^2\equiv\mathbf{j}_{i,\mathbf{x}}^\text{rot}\cdot\mathbf{j}_{i,\mathbf{x}}^\text{rot}$. Note that by virtue of the CL procedure, the latter quantity will be in general complex in single snapshots, while in the long-time average it must become real and positive. In \Fig{Ytilde} we show the real part of $\mathbf{j}_{i,\mathbf{x}}^\text{rot}\cdot\mathbf{j}_{i,\mathbf{x}}^\text{rot}$, multiplied by ${m^2\xi_\mathrm{h}^2}[{2\pi \ln(L/\xi_\text{h})\langle\rho\rangle^2}]^{-1}$.
%

\begin{figure}
	\includegraphics[width=0.9\columnwidth]{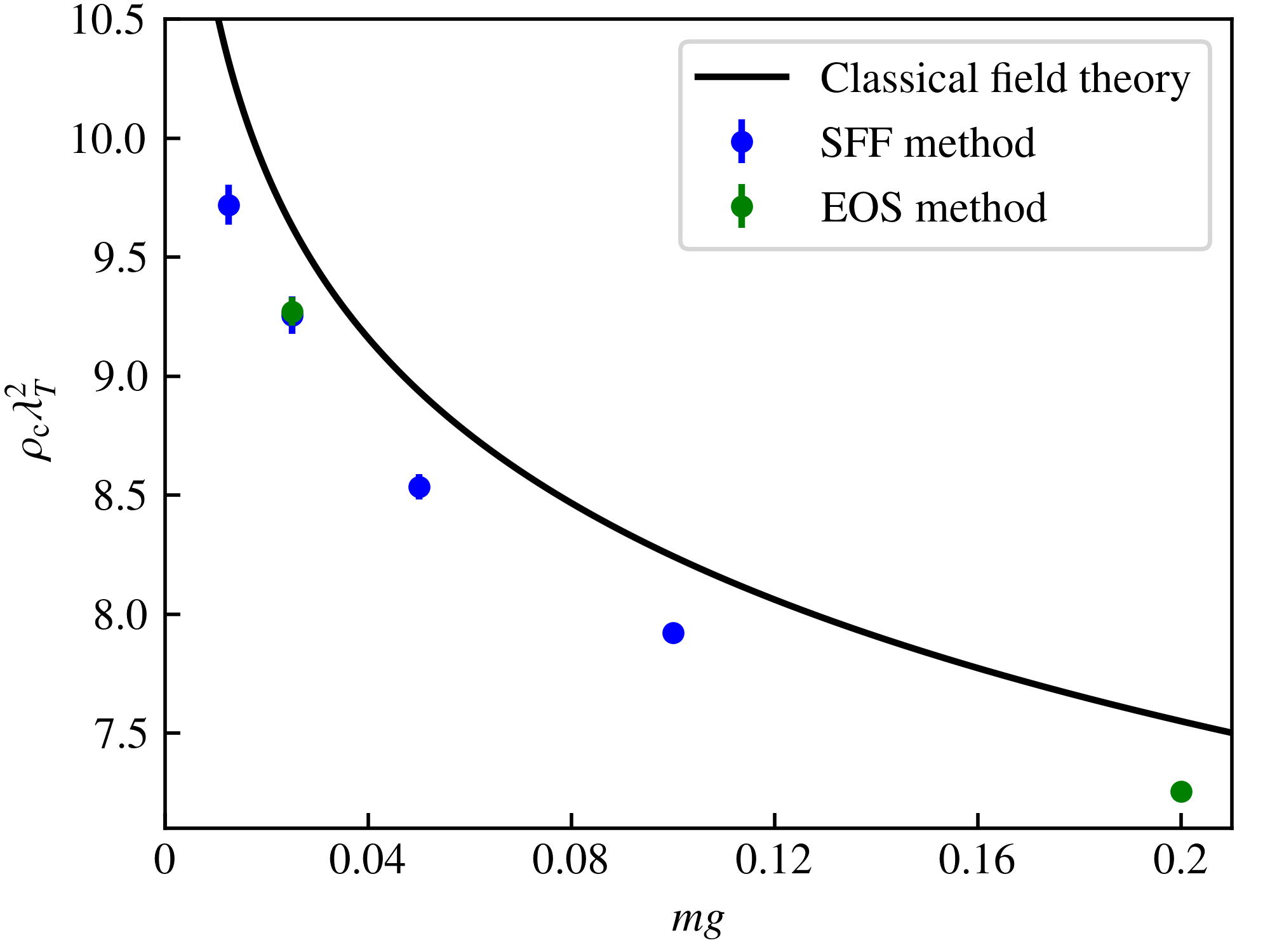}
	
	\caption{Critical density $\rho_\text{c}$ as a function of the coupling $mg$ computed by employing the Nelson criterion for the superfluid fraction (SFF), as described in \Sect{critdens} (blue points), in comparison to the results from the rescaling of the equation of state (green points). Since the latter yields only differences of critical densities, we chose the critical density at $mg=0.1$ to be equal to the result from the superfluid fraction method. The black line represents \Eq{critical_dense} with constant $\zeta_{\rho}=380$ obtained from the classical simulation of \cite{Prokofev2001a.PRL87.270402}.}
	\label{fig:rhocfromscaling}
\end{figure}

\section{\label{app:regfals} Finding the critical chemical potential with the secant algorithm}
We determined the chemical potential at the BKT transition point, i.e., the value $\mu_\mathrm{c}$, for which the Nelson criterion \eq{nelson} is fulfilled, by means of a secant algorithm, which converges faster than the common bisection algorithm. To this end, we start from two initial guesses for the critical chemical potential, $\mu_0$ and $\mu_1$, and determine subsequent estimates as 
\begin{align}
	\mu_{i+2}
	=\frac{\mu_{i}\tilde{\rho}_\mathrm{s}(\mu_{i+1})-\mu_{i+1}\tilde{\rho}_\mathrm{s}(\mu_{i})}
		{\tilde{\rho}_\mathrm{s}(\mu_{i+1})-\tilde{\rho}_\mathrm{s}(\mu_{i})}
	\,,
\end{align}
where $\tilde{\rho}_\mathrm{s}\equiv \rho_\mathrm{s}-\rho_\mathrm{s,c}$, and the critical superfluid density $\rho_\mathrm{s,c}$ is defined by \eq{nelson}. 
As initial guesses we chose $\mu_0=1.05\bar{\mu}$ and $\mu_1=0.95\bar{\mu}$ with $\bar{\mu}$ the estimate for the critical chemical potential from \cite{Prokofev2001a.PRL87.270402}, determined by \eq{critical_mu}. 
We stop the iteration once $\tilde{\rho}_\mathrm{s}$ is indistinguishable from $0$ within the statistical errors. This procedure is illustrated in \Fig{secant} (upper panel). 
As one can see, the convergence is extremely fast. 
In fact, we rarely had to perform more than 4 to 5 simulations for a given coupling and lattice size. 
Sometimes it occurred, however, that, due to statistical fluctuations, $\tilde{\rho}_\mathrm{s}(\mu_{i+1})$ and $\tilde{\rho}_\mathrm{s}(\mu_{i})$ resulted to be very close in magnitude, which could give rise to a divergence of the secant algorithm. 
In such cases we replaced $\tilde{\rho}_\mathrm{s}(\mu_{i})$ by hand by a suitable $\tilde{\rho}_\mathrm{s}(\mu_{j})$, with $j<i$, and thereafter continued the algorithm in the normal way.

\section{\label{app:critdens} Critical density from scaling}
As an application of the CL algorithm for determining the critical density for couplings $mg\gtrsim0.1$ was found to be increasingly impeded by runaway trajectories, we could not directly determine  $\rho_\mathrm{c}$ for the largest coupling $mg=0.2$ 
Nonetheless, as we could determine the equation of state further away from criticality, cf.~\Fig{eos}, where it was found to approximately scale according to \Eq{eosuniversal}, this allows us, under certain assumptions, to infer the critical density also for $mg=0.2$. 
In \Fig{rhocfromscaling}, we show the critical density, for $mg=0.025$ and $mg=0.2$ (green data points), as obtained by rescaling the equation of state for these couplings to the one for $mg=0.1$ and thus inferring it from the critical density $\lambda_T^2\rho_\mathrm{c}(mg=0.1)=7.921\pm0.032$.
The error bar is obtained as a combination of the error on the value for $mg=0.1$ and the fitting error obtained as the mean-square deviation of the rescaled curves from \Fig{eos}. 
For comparison, the directly determined values from \Fig{rhoc} are included as blue data points.

\section{\label{app:classsim} Classical field theory simulations}
\begin{figure}[t]
	\includegraphics[width=0.9\columnwidth]{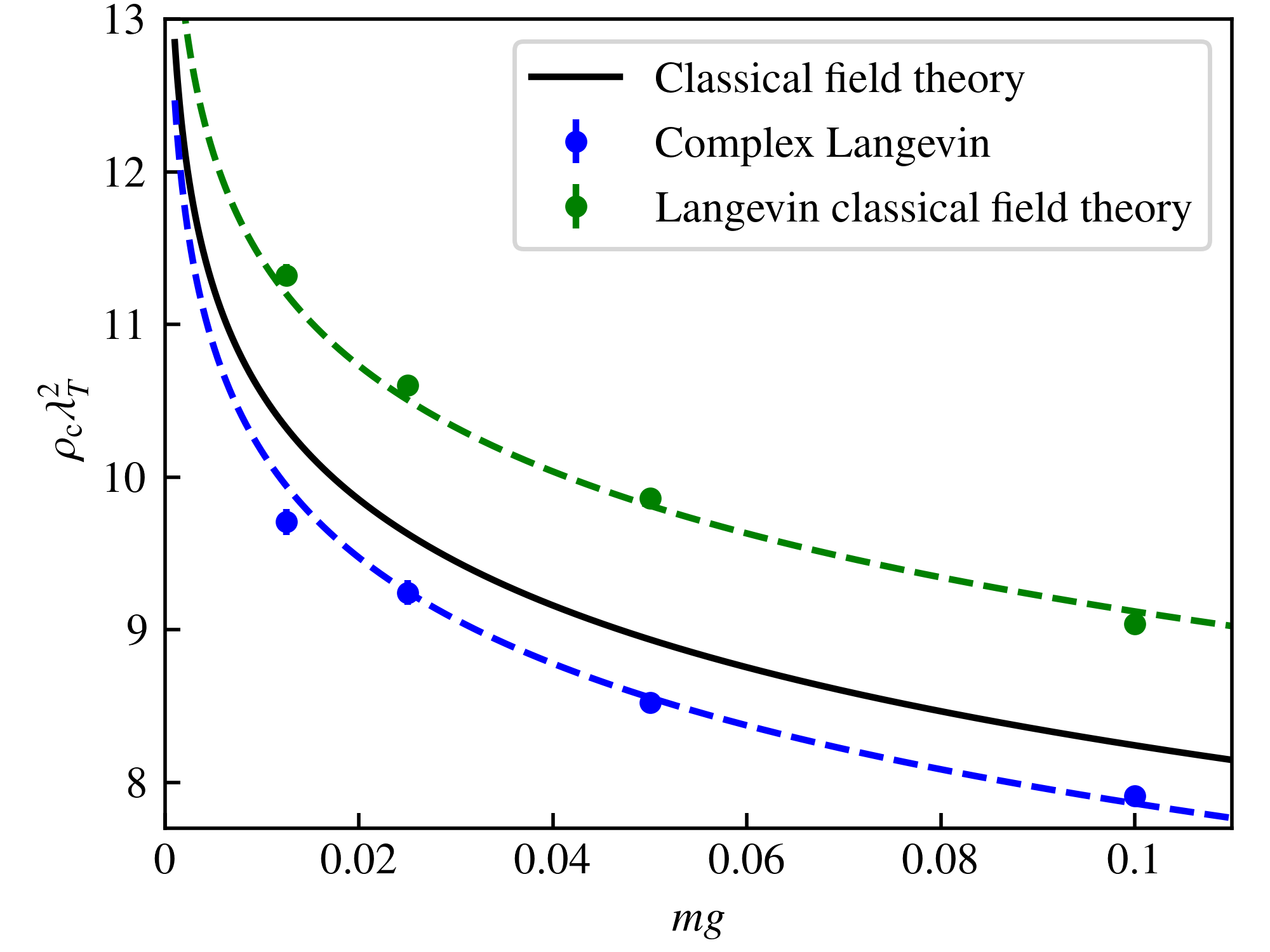}
	
	\caption{Critical density $\rho_\text{c}$ as a function of the coupling $mg$ computed by Langevin simulations of the classical field theory, i.e. setting $N_\tau=1$ (green points) in comparison with the full quantum simulations (blue points) and the result from the classical field theory simulation of \cite{Prokofev2001a.PRL87.270402}. The critical densities from the classical Langevin simulation are substantially higher than those from the full quantum simulation, and they result as higher than those from the classical field theory simulation \cite{Prokofev2001a.PRL87.270402}.}
	\label{fig:rhoclassical}
\end{figure}
\begin{figure}[t]
	\includegraphics[width=0.9\columnwidth]{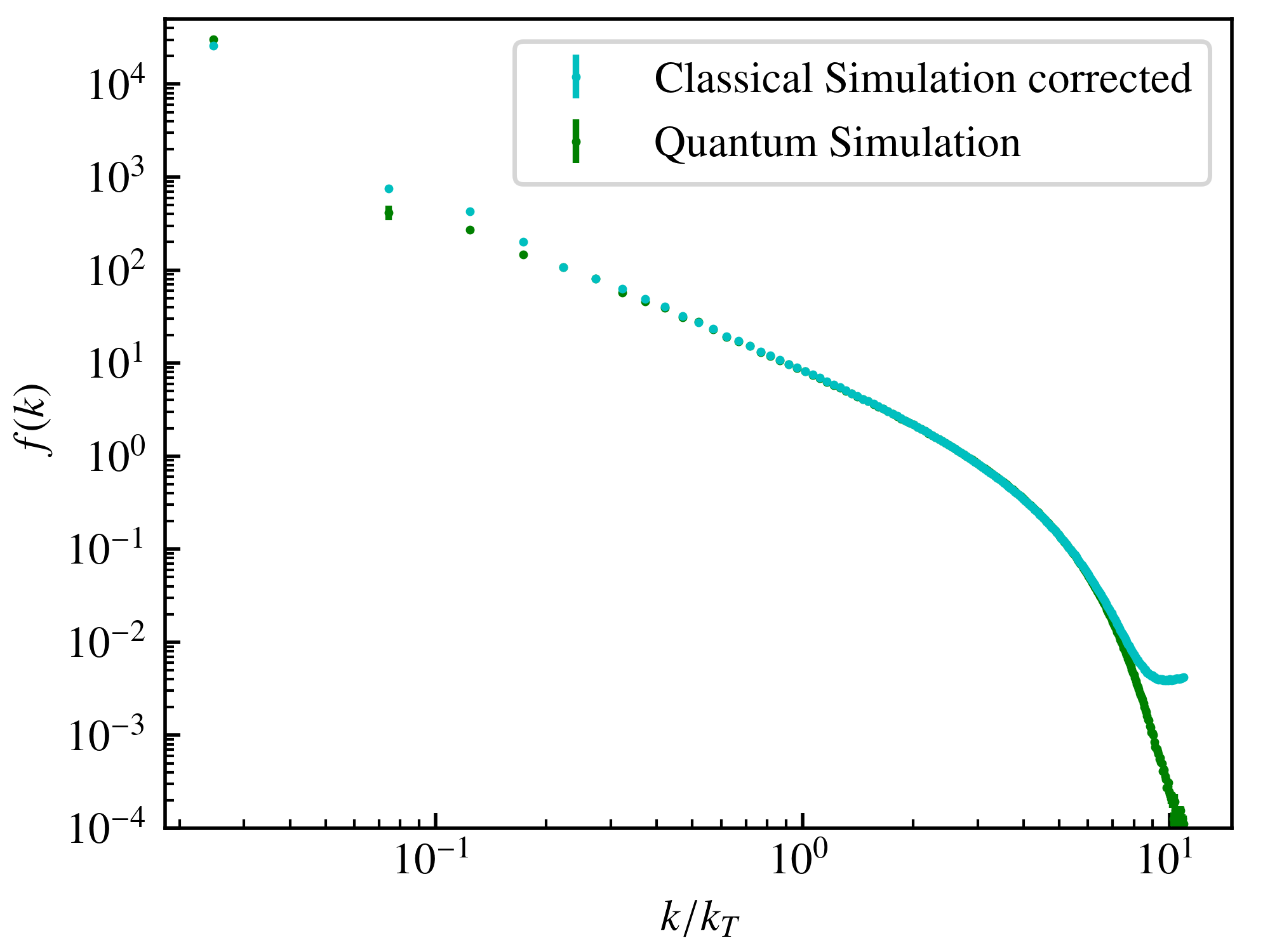}
	
	\caption{Comparison of the momentum spectrum from the full quantum simulation (blue points) and the classical simulation (green points) for $mg=0.1$ in the critical region, where the latter has been corrected in the UV by subtracting a Rayleigh-Jeans distribution and adding a Bose-Einstein distribution at zero chemical potential. 
	Note that the zero mode has been shifted to a finite $k$ value in order to make it visible on the double-logarithmic scale. 
	The chemical potential $\mu$ of the quantum simulation is matched to the one of the classical simulation, $\mu^\text{class}$, according to $\mu=\mu^\text{class}-2g\Delta\rho$, \Eq{mumatch}. 
	Apart from small deviations at very high momenta, the spectra agree well over almost the entire momentum range. However, deviations are visible in the strongly correlated IR-modes, with the quantum system being farther in the superfluid phase than the classical one. 
	The corresponding density of the quantum system results only by approximately $2\%$ larger than the corrected density of the classical system. }
	\label{fig:classicalspectrum}
\end{figure}
Setting the number of lattice points in imaginary direction to one,  $N_\tau=1$, allows us to easily turn off quantum effects and to simulate a purely classical field theory. In this appendix, we briefly discuss the results of such simulations for comparison.

For such a theory, occupation numbers will follow a Rayleigh-Jeans law in the UV, $f(k)\sim \left(k^2/2m-\mu\right)^{-1}$, such that most quantities such as density or kinetic energy suffer from a UV divergence that is absent in the full quantum theory. In order to regulate said divergence, we follow the procedure of \cite{Prokofev2001a.PRL87.270402} and correct the densities and chemical potentials within classical field theory, $\rho^\text{class}$ and $\mu^\text{class}$, by subtracting the (non-interacting) Rayleigh-Jeans distribution and adding the (non-interacting) Bose-Einstein distribution, i.e.~we obtain the density and chemical potential $\rho$ and $\mu$ in the full quantum theory as
\begin{align}
	\label{eq:rhomatch}
	\rho&=\rho^\text{class}+\Delta\rho
	\,,\\
	\label{eq:mumatch}
	\mu&=\mu^\text{class}-2g\Delta\rho
	\,,
\end{align}
with 
\begin{align}
	\Delta\rho
	&\equiv\frac{1}{L^2}\sum_\mathbf{p}
	\left[\frac{1}{\exp(\beta{\mathbf{p}^2}/{2m})-1}-\frac{2m}{\beta{\mathbf{p}^2}}\right]\,.
\end{align}
For the lattice spacing employed here, $\lambda_T=3.17\,a_\mathrm{s}$, this results in $\Delta \rho = 0.227\,a_\mathrm{s}^{-2}$. 
In \cite{Prokofev2001a.PRL87.270402}, the superfluid density was estimated by the number of windings in the systems, which are directly accessible within the employed worm algorithm. 
Since we have no access to winding numbers, we need, as in the quantum simulations in the main text, to resort to \Eq{rhos}, which requires to deal with the UV divergence of $\langle\mathbf{P}^2\rangle$ that is absent in the quantum simulation. 
We here follow the same strategy as for the density and subtract the value of $\langle\mathbf{P}^2\rangle$ in a non-interacting classical field theory and add its value in a non-interacting quantum field theory. The latter reads
\begin{align}
	\langle\mathbf{P}^2\rangle^\text{free}
	&=\sum_\mathbf{p} \left\{\frac{1}{\exp(\beta\frac{\mathbf{p}^2}{2m})-1}
	+\frac{1}{\left[\exp(\beta\frac{\mathbf{p}^2}{2m})-1\right]^2}\right\}\mathbf{p}^2\,,
\end{align}
such that we obtain the true $\langle\mathbf{P}^2\rangle$ from $\langle\mathbf{P}^2\rangle^\text{class}$ as 
\begin{align}
	\langle\mathbf{P}^2\rangle
	=\langle\mathbf{P}^2\rangle^\text{class}+\Delta\langle\mathbf{P}^2\rangle
	\,,
\end{align}
with 
\begin{align}
	\label{eq:deltaP2}
	\nonumber\Delta\langle\mathbf{P}^2\rangle
	\equiv \sum_\mathbf{p} &\left\{
	\frac{1}{\exp(\beta{\mathbf{p}^2}/{2m})-1}
	+\frac{1}{\left[\exp(\beta{\mathbf{p}^2}/{2m})-1\right]^2}\right.
	\\
	&\left.-\ \frac{2m}{\beta{\mathbf{p}^2}}
	-\left(\frac{2m}{\beta{\mathbf{p}^2}}\right)^2\right\}\mathbf{p}^2
	\,,
\end{align}
which for our lattice gives $\Delta\langle\mathbf{P}^2\rangle/L^2=1.41\,a_\mathrm{s}^{-4}$.

In this way, we have repeated the extraction of the critical densities described in the main text for the classical field theory, results of which are shown in \Fig{rhoclassical}. 
The critical densities are by around $15\%$ larger than in the full quantum simulation and larger than the results from \cite{Prokofev2001a.PRL87.270402}, which could indicate that the different classical field theory simulations are not entirely compatible with each other or some residual UV cutoff dependence remains.

In order to shed more light at the discrepancy between classical and quantum simulations, we compare, in \Fig{classicalspectrum}, the momentum spectra obtained from the classical and quantum simulations, for $mg=0.1$, with the chemical potentials matched according to $\mu=\mu^\text{class}-2g\Delta\rho$. 
After subtracting the free Rayleigh-Jeans distribution and adding the Bose-Einstein distribution, the two spectra agree with each other for $k>k_\mathrm{c}$, apart from a small deviation in the far UV that is due to the Rayleigh-Jeans fall-off being slightly weaker than $k^{-2}$ and that, as we checked, has little effect on both, the particle number and the superfluid density. 
For the strongly correlated IR modes, $k<k_\mathrm{c}$, however, one observes a substantial deviation, with the quantum system being already farther in the superfluid phase. 
This suggests that, while the effect of the density bias $\Delta \rho$ of the classical simulation can be accounted for in the description of the modes $k>k_\mathrm{c}$ by a shift of the effective chemical potential by $2g\Delta\rho$, its effect on the effective chemical potential governing the IR is more involved. 
The latter could be expected in view of the fact that at the transition point, the system is already substantially condensed, while the approximation $\mu_\text{eff}=\mu-2g\rho$ derives from the Hartree-Fock substitution of the interaction term, which is valid only in the non-condensed phase. 

\end{appendix}

%

\end{document}